\documentclass[a4paper, nofootinbib,article,11pt,longbibliography,superscriptaddress]{revtex4-1}
\usepackage{dcolumn}   
\usepackage{bm}        
\usepackage{amssymb}   
\usepackage{xcolor}
\usepackage[normalem]{ulem}
\usepackage{amsmath}
\usepackage{mathtools}
\usepackage[english]{babel}
\usepackage{accents}
\usepackage{natbib}
\usepackage{enumitem}
\usepackage[bottom]{footmisc}
\usepackage[section]{placeins}
\usepackage{dsfont}
\usepackage{natbib}
\usepackage{relsize}
\usepackage{etoolbox}
\patchcmd{\section}
  {\centering}
  {\raggedright}
  {}
  {}
\patchcmd{\subsection}
  {\centering}
  {\raggedright}
  {}
  {}
  
  \usepackage{mciteplus}
    \usepackage{hyperref}

\newcommand{\be}{\begin{equation}}
\newcommand{\ee}{\end{equation}}
\newcommand{\eps}{\varepsilon}

\newcommand{\Hdg}{\rm\scriptscriptstyle H}
\newcommand{\e}{\mbox{e}}
\newcommand{\de}{\mbox{d}}
\newcommand{\Tr}{\mbox{Tr}}
\newcommand{\spl}{\be\begin{split}}
\newcommand{\pa}{\partial}
\newcommand{\pha}{\phantom{a}}
\newcommand{\R}{\mathbb{R}}
\newcommand{\C}{\mathbb{C}}
\newcommand{\eqn}[1]{(\ref{#1})}

\def\beqa{\begin{eqnarray}}
\def\eeqa{\end{eqnarray}}
\def\bean{\begin{eqnarray*}}
\def\eean{\end{eqnarray*}}

\newcommand{\castellani}{castellani_A,*castellani_B}
\newcommand{\bimonteetal}{bimonte_A,*bimonte_B,*bimonte_C}
\newcommand{\bimetricCosmology}{Akrami:2015qga,*Akrami:2013ffa,*Konnig:2014xva,*Magueijo:2008sx,*vonStrauss:2011mq}
\newcommand{\bimetricReview}{deRham:2014zqa,*Schmidt-May:2015vnx}
\newcommand{\metricAffine}{Hehl:1994ue,*Sotiriou:2006qn,*Vitagliano:2010sr,*Capozziello:2007tj}
\newcommand{\nonmetricity}{Krasnov:2007ei,*Bengtsson:2007zx,*Latorre:2017uve}
\newcommand{\weyl}{Smolin:1979uz,*deCesare:2016mml}
\newcommand{\torsion}{Hehl:1976kj,*Hammond:2002rm,*Shapiro:2001rz}
\newcommand{\barberoImmirziField}{Torres-Gomez,*Calcagni:2009xz,*Mercuri:2009zt}
\newcommand{\doubleFieldTheory}{Aldazabal:2013sca,*Berman:2014jba,*Hull:2009mi}
\newcommand{\ncDFT}{Blumenhagen:2013zpa,*Mylonas:2014aga,*Aschieri:2017sug,*Marotta:2018swj}
\newcommand{\selfdualPalatini}{Jacobson:1988yy,*Samuel:1987td}
\newcommand{\selfdualMore}{Ashtekar:1991hf,*Ashtekar:1988sw}
\newcommand{\lqcWilsonEwing}{wilson-ewing1,*wilson-ewing2}
\newcommand{\aschierietal}{aschieri_fermions_SW,*Aschieri:2012in}
\newcommand{\LQG}{Ashtekar:2004eh,*thiemann,*Rovelli:2004tv}
\newcommand{\Doplicher}{Doplicher:1994zv,*Doplicher:1994tu}

\mciteSetSublistMode{f}

\begin{document}

\begin{flushleft}
KCL-PH-TH/2018-24
\end{flushleft}

\title{Noncommutative gravity with self-dual variables}

\author{Marco de Cesare}
\email{marco.de\_cesare@unb.ca}
\affiliation{Department of Mathematics and Statistics, University of New Brunswick, Fredericton, NB, Canada E3B 5A3}
  \author{Mairi Sakellariadou}
\email{mairi.sakellariadou@kcl.ac.uk}
\affiliation{Department of Physics, King's College London, Strand WC2R 2LS, London, United Kingdom}
  \author{Patrizia Vitale}
\email{patrizia.vitale@na.infn.it}
\affiliation{Dipartimento di Fisica ``E. Pancini'' Universit\`a di Napoli Federico II, Complesso Universitario di Monte S.~Angelo, via Cintia, 80126 Naples, Italy}
\affiliation{INFN, Sezione di Napoli, Complesso Universitario di Monte S.~Angelo, via Cintia, 80126 Naples, Italy}

  \begin{abstract}
{We build a noncommutative extension of Palatini-Holst theory on a twist-deformed spacetime, generalizing a model that has been previously proposed by Aschieri and Castellani. The twist deformation entails an enlargement of the gauge group, and leads to the introduction of new gravitational degrees of freedom. In particular, the tetrad degrees of freedom must be doubled, thus leading to a bitetrad theory of gravity. The model is shown to exhibit new duality symmetries.
The introduction of the Holst term leads to a dramatic simplification of the dynamics, which is achieved when the Barbero-Immirzi parameter takes the value $\beta=-i$, corresponding to a self-dual action. We study in detail the commutative limit of the model, focusing in particular on the role of torsion and non-metricity. The effects of spacetime noncommutativity are taken into account perturbatively, and are computed explicitly in a simple example. Connections with bimetric theories and the role of local conformal invariance in the commutative limit are also explored.
}

\end{abstract}

\maketitle


\tableofcontents
\section*{Introduction}
Spacetime noncommutativity may represent one of the key features characterizing the geometric structure of spacetime at the Planck scale, thus marking a radical departure from the standard description of spacetime as a Riemannian manifold. The need to go beyond the classical concept of a spacetime manifold is shared by several different approaches to quantum gravity \cite{\LQG,Oriti:2011jm,Perez:2012wv,Ambjorn:2012jv,Dowker:aza,\Doplicher}. From a classical point of view, the extension of the framework of Riemannian geometry has been considered  in some classes of modified gravity theories \cite{\torsion,\metricAffine,\weyl}. Noncommutative gravity models are being investigated since many years and from several perspectives \cite{chamsed,aschieri,\aschierietal,stein,dobrski,buric,chamsed2,aschieri_diffeo_twist,estrada,koba,banerjee,vassil,mukher,\castellani,\bimonteetal}. The crucial step is to build a dynamical theory of gravity which is consistent with the particular noncommutative structure one assumes, and which is able to recover general relativity in the regime where it has been tested.

In the present paper we elaborate on a  noncommutative generalization of the Palatini action for general relativity which has been originally proposed in \cite{aschieri} and further developed in \cite{\aschierietal}. In the following, we consider the  full Palatini-Holst action, within the same geometric  framework. The noncommutative structure is obtained via a twist-deformation of the differential geometry \cite{aschieri_diffeo_twist,aschieri_stardiffeo,ALV}, which   allows  to build  from first principles a modified theory  of gravity having Planck scale modifications naturally built-in.  
There are two independent sources of new physical effects that may arise in such theories. 
Firstly, 
there are correction terms that become relevant close to the noncommutativity scale; in gravity models based on a twist-deformation, such as those in Refs.~\cite{aschieri,aschieri_dynamical_NC,aschieri_NCgaugefields,Aschieri:2014xka}
 and the one considered in this work, 
such corrections are readily obtained from an asymptotic expansion of the twist operator and lead to higher-derivative interactions.
Secondly, the commutative limit will be in general a modified theory of gravity, which extends general relativity with new degrees of freedom and extra terms in the action. This is indeed the case for the model presented in this work, where the introduction of extra fields is required for the consistency of the noncommutative theory. 
Such degrees of freedom survive in the commutative limit, and may thus have an impact on the behaviour of the gravitational interaction on large scales.

Following \cite{aschieri}, the theory we consider is formulated as a gauge model for a suitable extension of the Lorentz group, with twist-deformed spacetime. Noncommutativity of the algebra of fields defined over spacetime is therefore achieved through a so-called twist operator $\mathcal{F}= \exp{\left(-\frac{i}{2}\theta^{\alpha\beta}X_\alpha\otimes X_\beta\right)}$, such that the pointwise product $f\cdot g\equiv \mu\circ(f\otimes g)$ is replaced by a star-product $f\star g= \mu\circ\mathcal{F}^{-1}(f\otimes g)$. The parameters $\theta^{\alpha\beta}$ determine the noncommutativity scale.

Due to noncommutativity, the gauge group must be centrally extended in order for the gauge parameters to close under the Lie bracket. This is typical for noncommutative gauge theories and represents a non-trivial consistency requirement, which constrains the symmetry of the theory.
As a result, the Lorentz group is here centrally extended to $\rm GL(2,\C)$. The ensuing action functional is shown to be invariant under diffeomorphisms as well as $\star$-diffeomorphisms.  Because of the enlarged symmetry, new degrees of freedom have to be added to the model. Specifically, the Lorentzian spin connection must be replaced by a $\frak{gl}(2,\C)$ gauge connection.
Similarly, the tetrad must be replaced by a bitetrad. In Ref.~\cite{aschieri} those extra degrees of freedom are assumed to vanish in the commutative limit, where the standard Palatini action is recovered; this is achieved by imposing extra constraints. In subsequent papers the problem of finding solutions has been further addressed, mainly through the Seiberg-Witten map, or by perturbative expansion in the noncommutativity parameters \cite{Aschieri:2009qh,aschieri_fermions_SW,gespo1,gespo2,Aschieri:2014xka,Dimitrijevic:2014iwa,marija} (see also \cite{GarciaCompean:2003pm,GarciaCompean:2002ss}). A generalization of the model entailing dynamical noncommutativity was studied in Ref.~\cite{aschieri_dynamical_NC}, where the vector fields $X_\alpha$ entering the definition of the twist are promoted to dynamical variables.

 Our model contains a new term, which is obtained as a  twist-deformation of the Holst term \cite{Holst}. The coupling associated to it is the (inverse) Barbero-Immirzi parameter.
 The undeformed limit of the Holst term is known to be  topological if there are no sources of torsion, in which case it does not affect the field equations. Nonetheless, such term becomes dynamically relevant when torsion is taken into account. In the commutative theory, torsion is indeed non-vanishing when spinor fields are coupled to the gravitational field\footnote{Minimally coupled fermions to gravity in the noncommutative Palatini theory were studied in Ref.~\cite{aschieri,aschieri_fermions_SW}.}; the Barbero-Immirzi parameter $\beta$ then determines the strength of an effective four-fermion interaction \cite{perez,freidel}. However, when there are no sources of torsion (e.g.\ in vacuo), the commutative theory is exactly equivalent to general relativity. This is not the case for the noncommutative extension of the model: in fact, as a consequence of the bimetric nature of the theory, torsion is in general non-vanishing even in vacuo.
A further extension of the Holst action, obtained by promoting the parameter $\beta$ to a dynamical field (scalarization) was first considered in Ref.~\cite{taveras}, where it was hinted that it could provide a natural mechanism for k-inflation. The consequences of a dynamical $\beta$ have been further examined in Ref.~\cite{\barberoImmirziField}, whereas the running of $\beta$ in the context of Asymptotic Safety was studied in Ref.~\cite{reuter}.

  The Barbero-Immirzi parameter $\beta$ is well-known for playing an important role in loop quantum gravity, where it enters the definition of the Ashtekar-Barbero variables \cite{ashtekar_newvar,barbero,immirzi}. In the early formulation of the theory it was assumed to take either of the two values $\beta=\mp i$.  Such values correspond, respectively, to self-dual and anti-self dual spin connections  \cite{ashtekar_newvar}. Self-dual variables have also been considered more recently in loop quantum cosmology models in Refs.~\cite{\lqcWilsonEwing}. 
  The Lagrangian formulation of the Palatini-Holst theory with self-dual variables was obtained in Refs.~\cite{\selfdualPalatini}. Such theory is inherently complex and suitable reality conditions must be imposed on solutions to the field equations. We shall see in this work that the noncommutative extension of the model is entirely analogous in this regard. More specifically, for $\beta=\mp i$ the noncommutative Palatini-Holst theory only depends on a self-dual (resp.\ anti self-dual) $\frak{gl}(2,\C)$ gauge connection. The notion of duality in this context is closely related to the chirality operator, and reduces to the standard Hodge dual for the Lorentz component of the gauge connection.
   The choice of self-dual variables turns out to be particularly convenient to study the dynamics in our model since it reduces the number of extra dynamical degrees of freedom, thus leading to a considerable simplification in the equations of motion.
In relation to the original model presented in Ref.~\cite{aschieri} a self-dual connection was already considered  in Ref.~\cite{gespo2}, although the self-duality request was limited to the Lorentz component of the gauge connection,   as a working assumption to find solutions to the equations of motion. This is to be contrasted with self-duality of the full $\frak{gl}(2,\C)$ gauge connection, which is naturally enforced in our model for $\beta=-i$.

  An interesting feature of noncommutative extensions of general relativity such as the one we consider here is the much greater richness of the underlying geometric structure, which survives in the commutative limit. In fact, as we will discuss more in detail later, such theories are generally bimetric, featuring both torsion and non-metricity.
 Ghost-free bimetric theories were originally constructed in \cite{Hassan:2011zd} as consistent non-linear theories of interacting spin-2 fields. They represent an extension of ghost-free massive gravity, first proposed in \cite{deRham:2010ik,deRham:2010kj} (see also the reviews~\cite{\bimetricReview}). The cosmological applications of bimetric gravity have been extensively studied, see e.g.\ Refs.~\cite{\bimetricCosmology}. 
  Bimetric (commutative) theories with gauge group $\rm GL(2, \C)$ have been previously  considered in Ref.~\cite{godlewski}. Such symmetry group implies in particular invariance under local Weyl rescalings; this feature is recovered in our model in the commutative limit.
The framework we consider is also a generalization of that of metric-affine theories \cite{\metricAffine}, where the metric and the affine connection (or, equivalently, the tetrad and the spin connection) are regarded as independent dynamical variables, thus allowing for non-vanishing torsion and non-metricity. In our case the independent dynamical variables are the bitetrad and the $\frak{gl}(2,\C)$ gauge connection. The physical properties and the possibility of detecting such departures from Riemannian geometry are discussed e.g.\ in Refs.~\cite{\torsion} for torsion and in Refs.~\cite{\nonmetricity} for non-metricity. Torsion is also particularly relevant for its potential role in Lorentz violation \cite{kostelecky}. 
A particular type of non-metric connection, namely the Weyl connection, can be used to formulate extended theories of gravity that are locally conformally invariant and without higher-derivatives \cite{\weyl}. This type of non-metricity is naturally realized in the model considered in this work; indeed, we will show that one of the components of the $\frak{gl}(2,\C)$ connection can be identified with the Weyl one-form.

Let us outline the main analysis and results of our study. We provide a detailed analysis of the dynamics and symmetries of the noncommutative Palatini-Holst theory in the pure gravity case. Fixing the value $\beta=-i$ for the Barbero-Immirzi parameter, we obtain the equations of motion for the model with self-dual variables. This choice leads to a dramatic simplification in the dynamics and reduces the number of extra degrees of freedom required by the noncommutative deformation. We  examine the commutative limit of the theory and solve the equations of motion for the gauge connection, showing that both torsion and non-metricity will be non-vanishing in general. This is essentially due to the fact that we are dealing with a bimetric theory. The non-metricity is of the Weyl type and is related to the extra component in the gauge connection. The gravitational field equations in this limit turn out to be equivalent to Einstein-Cartan equations, with the field strength of the Weyl vector acting as a source for torsion.
We perform an asymptotic expansion of the equations of motion up to second order in the deformation parameters $\theta^{\alpha\beta}$. In order to better elucidate the physical implications of the noncommutative corrections, we adopt a perturbative scheme to solve the equations. We then determine the perturbative corrections to a very simple solution of the model, in which the two tetrads are related by a constant scale transformation. In this particular case, the commutative limit yields vanishing torsion and vanishing field strength for the Weyl vector. However, our results show that the noncommutative corrections would in general give rise to a non-vanishing torsion to first order in $\theta^{\alpha\beta}$.

The model exhibits three different kinds of discrete symmetries (dualities), which are essentially due to its chiral nature. One of them is an obvious generalization of the Hodge duality of the Palatini-Holst action. The remaining two dualities are new features of this model, and crucially hinge on the doubling of the tetrad degrees of freedom. This is particularly interesting in relation to the general framework of double field theory \cite{\doubleFieldTheory}, and may pave the way for further investigations of non-geometric structures in noncommutative gravity models \cite{\ncDFT}.

The remainder of the paper is organized as follows. In Section \ref{holst}, we review the formulation of Palatini-Holst action in the tetrad formalism, stressing the role of the Lorentz group as an internal gauge symmetry. We discuss the self-dual and anti self-dual theories obtained for $\beta=- i,\,+i$~, respectively. We start in Section \ref{Section:NCextension} by reviewing deformed gauge transformations and discuss the new gravitational degrees of freedom. Then we adapt the formalism introduced in Ref.~\cite{aschieri} to the Palatini-Holst action. We establish a relation with bimetric theories of gravity and examine the possibility of introducing bitetrad interactions. Higher-order invariants are also discussed. Section \ref{noncomm} is devoted to the self-dual noncommutative case. It is shown that some well-known results concerning the decomposition of a gauge connection into its self-dual and anti self-dual components continue to hold in the noncommutative case, thereby showing that the anti-self dual component  of the gauge connection is projected out for $\beta=-i$. In Section \ref{sec:eom}, we derive the equations of motion from the self-dual action. In Section \ref{symm}, we analyse the symmetries of the model. We show that, besides being invariant under diffeomorphisms, $\star$-diffeomorphisms, and deformed $\rm GL(2,\C)$ gauge transformations, the noncommutative Palatini-Holst theory turns out to be invariant also under duality transformations; we identify three different types of such dualities. Our focus, in Section \ref{commlim}, is on the dynamics of the self-dual theory in the commutative limit. We solve the equations of motion for the gauge connection, showing that torsion will be in general non-vanishing. We show that the gravitational field equations obtained from the model can be recast in a form that is similar to Einstein-Cartan theory. The extra component in the $\frak{gl}(2,\C)$ gauge connection is identified with the Weyl one-form and acts as a source for torsion. In Section \ref{thetaexp}, we expand the noncommutative equations of motion to second order in the deformation parameters. In Section \ref{pertsol}, we adopt a perturbative scheme to solve the equations of motion for the gauge connection in the case of conformally related tetrads, showing that noncommutativity is a source of torsion to first order in $\theta^{\alpha\beta}$. In Section \ref{interact} we discuss further extensions of the model. Finally, in Section \ref{concl}, we highlight our concluding remarks. Five appendices containing technical details and review material complete the paper.

\

\noindent{\sl{  Notation and Conventions.}}
We use capital Latin letters to denote internal Lorentz indices. Small Latin letters denote spacetime tensor indices in Penrose abstract index notation.  
Greek indices will be used to label non-dynamical background structures (e.g.~the fields $X^\alpha$ and the parameters $\theta^{\alpha\beta}$) introduced by the twist-deformation.
Moreover, given a Lie group $G$,  we shall denote with  $\mathfrak{g}$ its Lie algebra. We will be often working with the universal covering ${\rm SL(2,\C)}$ of the Lorentz group $\rm SO(1,3)$, as well as with $\rm SO(1,3)$ itself; to keep the notation as simple as possible we shall identify the two.
 The hatted objects $\widehat G$
and $\widehat{\mathfrak{g}}$ will indicate the (infinite dimensional) group and the algebra of gauge
transformations, respectively; e.g. $\widehat{\rm G}= \{\Lambda: M\rightarrow \rm G\}$, where $M$ indicates the spacetime manifold and $\Lambda$ is a smooth map. Our metric convention for the Minkowski metric will be $\eta={\rm diag}(+,-,-,-)$. The conventions adopted for the Riemann and the torsion tensors are given in Appendix~\ref{Appendix:Torsion}.
\section{The Palatini-Holst action}\label{holst}
It is well known that the Einstein-Hilbert action for general relativity can be given a first order formulation, the Palatini action, 
where there is an internal $\widehat{\rm SO(1,3)}$ gauge symmetry. The internal space is specified by the tetrad frame $e^I_a$.
The theory is clearly also invariant under the group of spacetime diffeomorphisms ${\rm Diff}(M)$, where $M$ is the spacetime manifold. 
The Palatini action can be 
generalized by supplementing a new term, known as the Holst term \cite{Holst}, which does not have any effect on the equations of motion as long as torsion vanishes. 
In fact, in the torsion-free case, such term plays for the classical theory an analogue role to topological contributions in Yang-Mills theories. However, the situation is different when torsion is non vanishing \cite{perez}.

Let us  review the Palatini-Holst theory in the commutative case.  The action reads as
\be\label{Eq:HolstAction}
S[e,\omega]=\frac{1}{16\pi G}P_{IJKL}\int  e^I\wedge e^J\wedge F^{KL}( \omega)~,
\ee
where
$e^I$ are tetrad one-forms, and the field strength $F^{IJ}$ is  the gauge curvature of the spin connection one-form $\omega$
\be\label{Eq:FieldStrengthDefinition}
F^{IJ}( \omega)=\de\omega^{IJ}+\omega^{IK}\wedge\omega_K^{~J}~.
\ee
$P_{IJKL}$ denotes the following tensor in the internal space
\be
P_{IJKL}=\frac{1}{2}\varepsilon_{IJKL}+\frac{1}{\beta}\delta_{IJKL}~,
\ee
where $\delta^{IJ}_{\pha\pha KL}$ is the identity on the space of rank-two antisymmetric tensors in the internal space
\be
\delta^{IJ}_{\pha\pha KL}=\delta^{I}_{\pha [K} \delta^{J}_{\pha L]}~.
\ee
The Palatini action is recovered from (\ref{Eq:HolstAction}) for  $1/\beta=0$. The $\beta$ parameter is complex-valued and is known as the Barbero-Immirzi parameter \cite{barbero, immirzi}.

When there are no sources of torsion (e.g., in the pure gravity case), the dynamics which follows from the Holst action is equivalent to the standard Palatini theory\footnote{This statement holds true only in the classical theory. In fact, in loop quantum gravity the Barbero-Immirzi parameter gives rise to a quantization ambiguity which affects the spectrum of geometric operators \cite{rovelli-thiemann}.}.
In fact, this is a consequence of the existence of a torsion topological invariant, namely the Nieh-Yan invariant \cite{Nieh_1982}
\be\label{Eq:NiehYan}
d(e^I\wedge T_I)=T^I\wedge T_I- e^I\wedge e^J\wedge F_{IJ} ~.
\ee
Thus, in the torsion-free case, the dynamics is not affected by the particular value of the Barbero-Immirzi parameter, which can be a priori an arbitrary complex number.

The action (\ref{Eq:HolstAction}) can also be recast in the following form, by explicitly writing  spacetime tensor indices\footnote{When the spin connection is on-shell, the action (\ref{Eq:CommutativeAction_TensorForm}) reduces to a functional of the metric $g_{ab}=\eta_{IJ}e^I_a e^J_b$ given by $S[g_{\mu\nu}]=-\frac{1}{16\pi G}\int \de^4 x \sqrt{-g}\, R$. The overall sign factor is chosen consistently with our convention for the metric signature.}
\be\label{Eq:CommutativeAction_TensorForm}
S[e,\omega]=\frac{1}{16\pi G}\left(\int \de^4 x\; e\,e^a_I e^b_J F^{IJ}_{ab}-\frac{1}{\beta}\int \de^4 x\; e\,e^a_I e^b_{J} *_{\Hdg} F^{IJ}_{ab}\right)~,
\ee
where $*_{\Hdg}$ denotes the Hodge star operator with respect to the internal space.
The Holst dual of the field strength is thus defined as
\be
*_{\Hdg} F^{IJ}=\frac{1}{2}\varepsilon^{IJ}_{\pha\pha KL} F^{KL}~.
\ee
The field strength of the gauge connection $\omega^{IJ}$ is related to the Riemann curvature of the corresponding affine connection via
\be
F^{IJ}_{ab}=R_{ab}^{\pha\pha IJ} ~.
\ee

In analogy with Yang-Mills gauge theories, we can reformulate the theory in index-free form by introducing Lie algebra valued differential forms. Following \cite{aschieri} we use the spinorial representation of the Lorentz group,  generated by the commutators of Dirac gamma matrices (see Appendix \ref{Clifford} for useful formulae concerning the Clifford algebra). Thus, the Lie algebra valued connection one-form reads as
\be\label{Eq:SpinConnectionDecompose}
{\bm \omega}=\frac{1}{2}\omega^{IJ}\Gamma_{IJ}~,
\ee
with $\Gamma_{IJ}=\frac{i}{4}[\gamma_I,\gamma_J]$, while its curvature two-form (field strenght) is given by
\be\label{Eq:FieldStrengthDecompose}
{\bm F}({\bm \omega})=\frac{1}{2}F^{IJ}( \omega)\Gamma_{IJ}=\de {\bm \omega}-i{\bm \omega}\wedge{\bm \omega}~.
\ee
 Moreover, tetrads are associated to  vector  valued one-forms \cite{aschieri} according to 
\be\label{Eq:TetradDecompose}
{\bm e}= e^I \gamma_I
\ee
The definition \eqn{Eq:FieldStrengthDecompose} of the field strength can be easily checked to be  completely equivalent to the standard one represented by  Eq.~(\ref{Eq:FieldStrengthDefinition}).
Then  the action can be recast in the form
\be\label{Eq:AC_FullAction_Commutative}
S[{\bm e},{\bm \omega}]=\frac{i}{32\pi G}\int \Tr\left[{\bm e}\wedge{\bm e}\wedge\left( *_{\Hdg}{\bm F}+\frac{1}{\beta}{\bm F}\right) \right]~
\ee
where   $\Tr[\dots ]$ represents the trace over the Lie algebra and use has been made of the following identity (cf.~Ref.~\cite{aschieri})
\be
*_{\Hdg}{\bm F}=-i{\bm F}\gamma_5~. \label{Hstardef}
\ee

The behaviour of dynamical fields under gauge transformations $\Lambda\in \widehat{{SO}(1,3)}$, representing Lorentz transformations in the internal space,  is the following
\begin{align}
{\bm e}&\rightarrow {\bm \Lambda}{\bm e}{\bm \Lambda}^{-1}\label{eq:Transf_Tetrad}\\
{\bm \omega}&\rightarrow{\bm \Lambda}{\bm \omega}{\bm \Lambda^{-1}}+i{\bm \Lambda}\de {\bm \Lambda^{-1}}\label{eq:Transf_Connection}\\   
{\bm F}&\rightarrow{\bm \Lambda}{\bm F}{\bm \Lambda^{-1}}\label{eq:Transf_FieldStrength}    
\end{align}
with the gauge transformation ${\bm \Lambda}$  given by
\be
{\bm \Lambda}(x) =\exp(- i {\bm \epsilon }(x)) ~~, \hspace{1em} \mbox{with} ~~ {\bm \epsilon}(x)=\frac{1}{2}\epsilon^{IJ}(x) \Gamma_{IJ}~.
\ee
The action \eqn{Eq:AC_FullAction_Commutative} is invariant under gauge transformations (\ref{eq:Transf_Tetrad})--(\ref{eq:Transf_FieldStrength}).
Expanding (\ref{eq:Transf_Tetrad})--(\ref{eq:Transf_FieldStrength}) to first order in the gauge parameter ${\bm \epsilon}$, we have the following infinitesimal form of gauge transformations
 \begin{align}
 {\bm e}&\rightarrow  {\bm e} +i [ {\bm e},{\bm \epsilon}]\label{eq:Transf_Tetrad_infinitesimal}\\
 {\bm \omega}&\rightarrow  {\bm \omega} -  \left( \de {\bm \epsilon} - i [ {\bm \omega},{\bm \epsilon}]\right) \\
 {\bm F}&\rightarrow  {\bm F} +i [ {\bm F},{\bm \epsilon}] ~.
 \end{align}

In our analysis of the dynamics of the noncommutative generalization of the Holst action in the following sections, a key role will be played by self-duality of the gauge connection.
The self-dual Palatini action was first studied in Ref.~\cite{\selfdualPalatini}, where it was used to derive Ashtekar's Hamiltonian formulation of general relativity \cite{ashtekar_newvar,ashtekar_newham} starting from a Lagrangian formulation (see also Refs.~\cite{peldan,\selfdualMore}).
In the context of noncommutative Palatini gravity, it has been already considered in Ref.~\cite{gespo2} as a simplifying assumption in order to find solutions for the model proposed in Ref.~\cite{aschieri}. In the Palatini-Holst theory, a theory of a self-dual gauge connection is naturally obtained by choosing a particular value for the Barbero-Immirzi parameter $\beta$.
In the next section, where the noncommutative extension of the Holst action is proposed, self-duality will result in a powerful simplification of the model, which will allow for a systematic study of the dynamics and solutions of the equations of motion. Let us therefore briefly review the definition and fix the notation.

A gauge connection is said to be self-dual (resp.~anti self-dual) if it is a solution to the eigenvalue equation\footnote{We recall that $*_{\Hdg}^2=-1$ on a four dimensional manifold with Lorentzian signature.}
\be\label{Eq:Self-Duality_EigenvalueEquation}
*_{\Hdg}{\bm \omega}_{\pm}=\pm i\, {\bm \omega}_{\pm}~,
\ee
with 
$
*_{\Hdg}{\bm \omega}\coloneqq -i{\bm \omega}\gamma_5 $. 
An analogous definition holds for for the field strength. 
In particular, we can decompose the gauge connection and the field strength  into their  self-dual and anti self-dual parts as
$
{\bm \omega}={\bm \omega}_{+}+{\bm \omega}_{-}~,~~
{\bm F}={\bm F}_{+}+{\bm F}_{-}~.
$
Thus, it is possible to show that the self-dual (resp. anti self-dual) part of the field strength is given by the field strength of the self-dual (resp. anti self-dual) part of the gauge connection
\be
{\bm F}_{\pm}({\bm \omega})=\de {\bm \omega}_\pm -i{\bm \omega}_\pm\wedge{\bm \omega}_\pm ={\bm F}_{\pm}({\bm \omega}_\pm). ~
\ee
The projectors onto the self-dual (anti self-dual) part are\footnote{The $\pm$ subscripts for the projectors here refer to the sign of the eigenvalues of the Hodge star operator $*_{\Hdg}$. With our conventions, the corresponding eigenvalues of the chirality operator $\gamma_5$ have the \emph{opposite sign}, see Appendix \ref{Clifford}.}
\be\label{Eq:ChiralProjectors_Define}
P_\pm=\frac{1}{2}(1\mp \gamma_5)~
\ee
with usual properties such as $(P_\pm)^2=P_\pm, ~~P_+ + P_- =\mathds{1}, ~
~
P_+ P_-=P_-P_+=0. ~$ 
Note that they coincide with the projectors onto the left-handed and the right-handed components of a Dirac spinor, respectively.
For instance, the self-dual part of the gauge connection is given by
\be\label{Eq:LeftHanded_SelfDual}
{\bm \omega}_+=P_+\, {\bm \omega}\, P_+ = (P_+)^2\, {\bm \omega}=\frac{1}{2}({\bm \omega} - \gamma_5 {\bm \omega})=\frac{1}{2}({\bm \omega}-i *_{\Hdg}{\bm \omega})~.
\ee
A similar relation holds for the anti self-dual connection. In the second equality we used the fact that $\gamma_5$ commutes with the generators of the Lorentz group. 
The components of ${\bm \omega}_\pm$ in the internal space read  
\be\label{Eq:SelfDualPartComponents}
\omega_\pm^{IJ}=\frac{1}{2}\left(\omega^{IJ} \mp \frac{i}{2}\varepsilon^{IJ}_{\pha\pha KL}\omega^{KL}\right)~.
\ee
Introducing the projectors onto the space of self-dual and anti self-dual two-forms (with regard to the internal space)
\be
p^{\pm \, IJ}_{\pha\pha\pha\, KL}=\frac{1}{2}\left(\delta^{IJ}_{\pha\pha KL}\mp \frac{i}{2}\varepsilon^{IJ}_{\pha\pha KL}\right)~,
\ee
equation (\ref{Eq:SelfDualPartComponents}) can be rewritten as
\be
\omega_\pm^{IJ}=p^{\pm \, IJ}_{\pha\pha\pha\, KL}\omega^{KL}~.
\ee
The tensors $p^{\pm \, IJ}_{\pha\pha\pha\, KL}$ can thus be regarded as the components counterpart of the operators $P_\pm$~. It is worth remarking the correspondence between self-duality and chirality, which is made transparent by the spinorial representation of the connection one-form (cf.~Ref.~\cite{jacobson-smolin_letter}).

Having introduced the definition and some basic properties of self-dual connections, we come back to the action (\ref{Eq:AC_FullAction_Commutative}). We observe that for $\beta=-i$ the anti self-dual component of the spin connection is projected out. Thus the action reduces to a functional of the tetrad and the self-dual connection only
\be\label{Eq:SelfDualAction}
S[{\bm e},{\bm \omega}_+]=
 \frac{i}{32\pi G}\int \Tr\left[{\bm e}\wedge{\bm e}\wedge\left( *_{\Hdg}{\bm F}+i{\bm F}\right) \right]= 
\frac{i}{16\pi G}\int \Tr\Big[{\bm e}\wedge{\bm e}\wedge \big(*_{\Hdg}{\bm F}_+({\bm \omega}_+) \big)\Big]~.
\ee
Note that this is formally equal to twice the Palatini action for a self-dual connection. Similarly, for $\beta=+i$, a theory of an anti self-dual connection is obtained. For definiteness, in the rest of this work we will only be concerned with self-dual connections, although it is evident that similar results can be obtained in the anti self-dual case.

Some additional remarks on the action (\ref{Eq:SelfDualAction}) are now in order. Although it is formally equivalent to the action for Palatini theory, the self-duality of the connection makes it inherently complex.
Thus, the solution space of the theory is much enlarged. Equivalence with the standard Palatini theory is then attained only after imposing suitable reality conditions. More precisely, one must require that that the tetrad $e^I$ be real, and that $\omega^{IJ}_{+}$ be the self-dual part of a real connection one-form
\be\label{Eq:RealityCondition_omega}
\omega^{IJ}_{+}=\frac{1}{2}\left(\omega^{IJ} - i *_{\Hdg} \omega^{IJ}\right)~.
\ee
Equation (\ref{Eq:RealityCondition_omega}) is equivalent to
\be
\Re\{\omega^{IJ}_{+}\}=*_{\Hdg} \left(\Im\{\omega^{IJ}_{+}\}\right) ~.
\ee

\section{Noncommutative extension}\label{Section:NCextension}
Noncommutative field theories are defined in terms of fields which are elements of a noncommutative algebra over space-time,  with a noncommutative, associative product. 
This is generally a deformation of a commutative algebra, which is recovered when the noncommutativity parameter is set to zero.  It is a general feature of noncommutative gauge theories that the Lie algebra of the structure group has to be extended in order for the algebra of gauge parameters to close. This is also the case for the Lorentz algebra of noncommutative gravity, as it has already been shown in \cite{aschieri}. Let us shortly review the derivation.

\subsection{Deformed gauge symmetry and symmetry enlargement}
Ordinary  gauge theories  with gauge group $\widehat G$
are modified by replacing the pointwise product of fields with a noncommutative  product indicated with $\star$. Here we use the noncommutative product defined in Eq.~\eqn{Eq:MoyalWeylProduct}, which is based on an Abelian twist. We refer to the  Appendix \ref{noncommapp} and to Ref.~\cite{aschieri} for more details.
The resulting field theories are invariant under the deformed gauge transformations
\be
\phi(x) \longrightarrow g_\star(x)\triangleright_\star \phi(x)=\exp_\star\left( i\,
\epsilon^i (x) T_i\right) \triangleright_\star \phi (x) ~, \label{gaugetransf}
\ee
 where $\phi(x)$ denotes a generic field in the theory, while $\triangleright_\star$ indicates generically the action of the group. 
 As an illustrative example, for non-Abelian gauge groups in their fundamental representation, the group action $\triangleright_\star$ entails a combination of matrix multiplication with the $\star$-product.  In equation (\ref{gaugetransf}) $T_i$ are the Lie algebra generators, $T_i \in \mathfrak{g}$,
and the gauge group elements $g_\star(x)$ are defined as star exponentials\footnote{For example,  the action on tetrad fields given in Eq. \eqn{eq:Transf_Tetrad} 
 becomes here
 \be
 {\bm e}(x) \longrightarrow {\bm\Lambda}_\star(x)\triangleright_\star  {\bm e}(x)={\bm \Lambda}_\star  \star {\bm e}\star {\bm \Lambda}_\star^{-1} ~(x)
\ee
with ${\bm \Lambda}_\star (x)= \exp_\star (- i {\bm \epsilon }(x))$ and 
${\bm \epsilon}(x)=\frac{1}{2}\epsilon^{IJ}(x) \Gamma_{IJ}$. }
\be
g_\star(x)=\exp_\star\left(i\, \epsilon (x)^i T_i\right)= 1+ i \epsilon^i
(x) T_i - \frac{1}{2}(\epsilon^i\star \epsilon^j)(x) T_i T_j + \ldots
\label{starexp}
\ee
At the infinitesimal level we have then\footnote{For notational simplicity, here we are using the same symbol $\triangleright_\star$ to denote the corresponding Lie algebra action induced by the group action defined in Eq.~(\ref{gaugetransf}).}
\be
\phi(x) \longrightarrow  \phi(x) +i (\epsilon\triangleright_\star \phi)(x) ~,
\label{infgaugetransf}
\ee
with
\be
(\epsilon\triangleright_\star\phi)(x)=\left(\epsilon^j\star (T_j\triangleright \phi)
\right)(x)
\ee
and $T_j$ is in the appropriate representation to  the field $\phi$.

The deformed Lie bracket reads
\be
[\epsilon_1,\epsilon_2]_\star(x)=(\epsilon_1 \star \epsilon_2) (x) - (\epsilon_2 \star \epsilon_1) (x) ~, \label{defolie}
\ee
Consistency of the theory demands that the algebra of infinitesimal gauge transformations must close under the $\star$-commutator defined by Eq.~(\ref{defolie}). 
However, it is evident that  in noncommutative field theory
 algebra closure is not guaranteed, since we have from the definition (\ref{defolie})
\be
\begin{split}
(\epsilon_1 \star \epsilon_2) (x) - (\epsilon_2 \star \epsilon_1) (x)=&
\frac{1}{2}\left((\epsilon_1^i \star \epsilon_2^j )(x)+(\epsilon_2^j\star
\epsilon_1^i)(x)\right) [ T_i, T_j]\nonumber\\
&+\frac{1}{2}\left ((\epsilon_1^i\star
\epsilon_2^j) (x)-(\epsilon_2^j\star \epsilon_1^i)(x)\right)
\{T_i,T_j\} ~,
\end{split}
\ee
which contains the anticommutator of the algebra generators. In fact, we note that in general the anticommutators of the generators of a given Lie algebra (for a given representation) are not elements of the algebra. Particular examples of Lie algebras which 
include the anticommutators are given by the Lie algebras of the unitary groups $U(N)$, for any $N$, in the adjoint and fundamental representations.
However, for the case under consideration, where we have chosen to represent the Lie algebra of the Lorentz group  in  the bispinor representation,  the anticommutators of the generators do not belong to the algebra, namely, the algebra closure is not attained.
Thus, we need to extend the $\mathfrak{sl}(2,\C)$ algebra in such a way as to include the anticommutators. Indeed, in the bispinor representation of $\mathfrak{sl}(2,\C)$, the anticommutators involve two further elements, namely $\mathds{1}_4$ and $\gamma_5$, which do not belong to the $\mathfrak{sl}(2,\C)$ algebra. In fact, we have for the case at hand (cfr.~Eq.~\eqn{Appendix:Eq:GeneratorsAntiCommutator})
\be
\{T_i,T_j\}\rightarrow \{ \Gamma_{IJ},\Gamma_{KL} \}= \eta_{I[K}\eta_{L]J}\mathds{1}+\frac{i}{2}\varepsilon_{IJKL}\gamma_5.
\ee
Thus, the sought for extended algebra is the $\mathfrak{gl}(2, \C)$ algebra in its bispinor representation.
$( \Gamma_{IJ}, \mathds{1}, \gamma_5 )$ is a set of generators of the (reducible) representation of $\mathfrak{gl}(2,\mathbb{C})$ on Dirac spinors, with $\mathds{1}$ and $ \gamma_5$ being central elements. Thus, we are led to enlarge the gauge algebra and consider more general infinitesimal gauge transformations of the form \cite{aschieri}
\be\label{Eq:Epsilon_GL2C_Define}
{\bm \epsilon}(x) =\frac{1}{2}\epsilon^{IJ}(x)\Gamma_{IJ}+ \epsilon(x) \mathds{1} + \tilde{\epsilon}(x) \gamma_5~.
\ee
\subsection{New gravitational degrees of freedom}
Next we study the transformation properties of the dynamical fields, namely the tetrad and the $\mathfrak{gl}(2,\mathbb{C})$ gauge connection, under a gauge transformation (cf.~Ref.~\cite{aschieri}). 
The generalization of  Eq.~(\ref{eq:Transf_Tetrad_infinitesimal}) to the noncommutative case is 
\be\label{Eq:GaugeTransfTetrad_Deformed_Definition}
\delta {\bm e}\coloneqq i [ {\bm e},{\bm \epsilon}]_{\star}~,
\ee
which is formally obtained from the corresponding transformation law in the commutative case by replacing the commutator with a $\star$-commutator (see Section \ref{Section:NCextension}). If  ${\bm e}$ is assumed to have the expression (\ref{Eq:TetradDecompose}) as in the commutative case, evaluation of the r.h.s.~of Eq.~(\ref{Eq:GaugeTransfTetrad_Deformed_Definition}), with $\bm \epsilon$ given by Eq.~(\ref{Eq:Epsilon_GL2C_Define}), gives
\be
\delta {\bm e}=-\frac{1}{2}\left(e^I \star \epsilon_I^{\pha J} +\epsilon_I^{\pha J} \star e^I\right)\gamma_J - \frac{1}{2}\left(e^I \star (*_{\Hdg} \epsilon)_I^{\pha J} -(*_{\Hdg} \epsilon)_I^{\pha J} \star e^I\right)\gamma_J \gamma_5 + (e^I \star \epsilon -\epsilon\star e^I)\gamma_I~.
\ee
We observe that terms proportional to $\gamma_I \gamma_5$ are generated under gauge transformations of the tetrad. Such terms cannot be expressed as a linear combination of Dirac $\gamma_I$ matrices. Thus, we are led to consider more general objects of the form
\be\label{Eq:BiTetrad_Define}
{\bm e} \coloneqq e^I \gamma_I+\tilde{e}^I \gamma_I\gamma_5
\ee
Such object belongs to a representation of $\mathfrak{ gl}(2,\mathbb{C})$, which can be split into two irreducible representations of the Lorentz algebra $\mathfrak{ sl}(2,\mathbb{C})$. In fact, the two terms in Eq.~(\ref{Eq:BiTetrad_Define}) define a polar and an axial vector, respectively
\begin{align}
{\bm e}_P &= e^I \gamma_I~\\
{\bm e}_A &= \tilde{e}^I \gamma_I\gamma_5~.
\end{align}
The generator $\gamma_5$ acts on ${\bm e}$ by exchanging its polar and axial components. Since $e^I$ and $\tilde{e}^I$ identify two a priori  independent sets of four differential forms, or tetrad frames, we will refer to ${\bm e}$ as the bitetrad one-form. It is straightforward to show that the definition of the bitetrad ${\bm e}$ given in Eq.~(\ref{Eq:BiTetrad_Define}) is consistent with $\mathfrak{gl}(2,\mathbb{C})$ $\star$-gauge symmetry, i.e.~the variation $\delta {\bm e}$ under a gauge transformation defined by Eq.~(\ref{Eq:GaugeTransfTetrad_Deformed_Definition}) is still of the form (\ref{Eq:BiTetrad_Define}).

The general relation between two independent anholonomic frames (not necessarily equivalent) is given by
\be
\tilde{e}^I=M^{I}_{\pha J}\, e^J~, \label{tile}
\ee
with ${M}^{I}_{\pha J}$ a local $\rm GL(4,\mathbb{R})$ transformation, that is $M^{I}_{\pha J}\in  \widehat {\rm GL(4,\R)}$. In the noncommutative case, this relation shall be replaced by
\be\label{Eq:GeneralStarDecompositionTetrads}
\tilde{e}^I={M_\star}^{I}_{ \; J}\star e^J ~ 
\ee
with ${M_\star}^{I}_{ \; J}$ defined as a star exponential of $\frak{gl}(4,\R)$ algebra generators, as in Eq.~\eqn{starexp}. 
Since in our model the gauge group is $\rm \widehat{ GL_\star(2,\mathbb{C})}$, we choose to restrict the freedom expressed by Eq.~(\ref{Eq:GeneralStarDecompositionTetrads}) and allow for $M^{I}_{\star \; J}$ just  
 to belong to  the subgroup $\rm \widehat{ GL_\star(2,\mathbb{C})}\subset \rm  \widehat {GL_\star(4,\R)}$. 
Such a choice is, in fact, the minimal one that is compatible with deformed gauge invariance. 
In the commutative limit, the assumption (\ref{Eq:GeneralStarDecompositionTetrads}) has a clear geometrical interpretation, meaning that the two frames are unsheared, i.e.~they are mapped into each other by means of the composition of local Lorentz transformations and Weyl rescalings. Therefore, in the commutative limit, such an assumption implies that the two frames refer to conformally equivalent metrics (see Section \ref{Sec:GaugeTransf}).

As a consequence of the enlargement of the gauge symmetry algebra to $\widehat{\mathfrak{ gl}_\star(2,\mathbb{C})}$, the field content of the theory is extended as follows:
\begin{align}
{\bm e} &= {\bm e}_P+{\bm e}_A=e^I \gamma_I+\tilde{e}^I \gamma_I\gamma_5~, \label{Eq:RepeatDefinition_Bitetrad} \\
{\bm \omega} &=\frac{1}{2}\omega^{IJ}\Gamma_{IJ}+\omega \mathbb{I}+\tilde{\omega}\gamma_5~.  \label{Eq:RepeatDefinition_Omega}
\end{align}
The ordinary spin connection of the commutative theory is given by the Lorentz part of the ${\mathfrak{ gl} (2,\mathbb{C})}$ connection ${\bm \omega}$.
The field strength of the gauge connection ${\bm \omega}$ is defined as
\be\label{Eq:Define_FieldStrength}
{\bm F}({\bm \omega})=\de {\bm \omega}-i{\bm \omega}\wedge_{\star}{\bm \omega}~
\ee
where the ordinary wedge product has been replaced by the twist-deformed wedge product, resulting from the composition with the twist operator (see Eq. \eqn{Eq:DeformedWedgeProd_Define}). We have then 
\be
{\bm F} =\frac{1}{2}F^{IJ}\Gamma_{IJ}+r\mathbb{I}+\tilde{r}\gamma_5~ \label{Eq:FieldStrenghtDecomposition}
\ee
with the  field strength components given by 
\begin{align}
F^{IJ} &=\de \omega^{IJ}+\frac{1}{2}\left(\omega^I_{\pha K} \wedge_{\star} \omega^{KJ}-\omega^J_{\pha K} \wedge_{\star} \omega^{KI}\right)-\frac{i}{2}(\omega^{IJ}\wedge_{\star}\omega+\omega\wedge_{\star}\omega^{IJ})+\nonumber\\
&\phantom{=}~~ \frac{1}{2}\left(*_{\Hdg} \omega^{IJ}\wedge_{\star}\tilde{\omega}+\tilde{\omega}\wedge_{\star}*_{\Hdg} \omega^{IJ}\right)  ~~~,\label{Eq:Eq:FieldStrenghtDecompositionExpanded1}   \\
r &=\de\omega-\frac{i}{8}(\omega^{IJ}\wedge_{\star} \omega_{IJ})-i(\omega\wedge_{\star}\omega+\tilde{\omega}\wedge_{\star}\tilde{\omega})  ~~~,\label{Eq:Eq:FieldStrenghtDecompositionExpanded2}   \\
\tilde{r} &= \de \tilde{\omega}+\frac{1}{8}\omega^{IJ}\wedge_{\star} *_{\Hdg} \omega^{IJ}-i(\omega\wedge_{\star}\tilde{\omega}+\tilde{\omega}\wedge_{\star}\omega)~~~.\label{Eq:Eq:FieldStrenghtDecompositionExpanded3}
\end{align}

\subsection{The action of internal symmetries on fields}\label{Sec:GaugeTransf}
Under an infinitesimal $\star$-deformed gauge transformation, the fields representing the basic dynamical variables of the theory transform as
 \begin{align}
 {\bm e}&\rightarrow  {\bm e} +i [ {\bm e},{\bm \epsilon}]_{\star}  \label{Eq:BiTetrad_GaugeTransform}\\
 {\bm \omega}&\rightarrow  {\bm \omega} -  \left( \de {\bm \epsilon} - i [ {\bm \omega},{\bm \epsilon}]_{\star}\right) \label{Eq:Connection_GaugeTransform}\\
  {\bm F}&\rightarrow  {\bm F} +i [ {\bm F},{\bm \epsilon}]_{\star}   \label{Eq:Curvature_GaugeTransform}
 \end{align}
 Equation~(\ref{Eq:Connection_GaugeTransform}) leads us to the following definition of the deformed covariant derivative of the gauge parameters ${\bm \epsilon}$
 \be
 D_{\bm \omega}^{\star} {\bm \epsilon} \coloneqq \de {\bm \epsilon} - i [ {\bm \omega},{\bm \epsilon}]_{\star}~.
 \ee

The physical meaning of the enlarged gauge symmetry becomes transparent in the commutative limit. To illustrate it, we consider infinitesimal transformations of the bitetrad given by Eq.~(\ref{Eq:BiTetrad_GaugeTransform}), which, in the commutative limit reduce to \eqn{eq:Transf_Tetrad_infinitesimal}, namely
\be\label{Eq:GaugeTransform_Bitetrad_CommutativeLimit}
\delta {\bm e}= i [ {\bm e},{\bm \epsilon}]~,
\ee
with ${\bm \epsilon}$ given by Eq.~(\ref{Eq:Epsilon_GL2C_Define}). The gauge group has two Abelian subgroups, generated, in the representation adopted here, by $\mathds{1}$ and $\gamma_5$. Clearly, Abelian gauge transformations generated by $\mathds{1}$ have no effect on the bitetrad. Computing the commutator in Eq.~(\ref{Eq:GaugeTransform_Bitetrad_CommutativeLimit}) we get
\be\label{Eq:GaugeTransform_Bitetrad_CommutativeLimit_Explicit}
\delta {\bm e}=\epsilon^I_{\pha J}e^J \gamma_I+\epsilon^I_{\pha J}\tilde{e}^J \gamma_I\gamma_5+2i\tilde{\epsilon}\left( e^I\gamma_I\gamma_5+ \tilde{e}^I\gamma_I \right)~.
\ee
Therefore, under an infinitesimal gauge transformation, the bitetrad components transform as
\begin{align}
\delta e^I =\epsilon^I_{\pha J}e^J+2i\tilde{\epsilon}\,\tilde{e}^I      ~, \label{Eq:GaugeTransform_Bitetrad_CommutativeLimit1}\\
\delta \tilde{e}^I =\epsilon^I_{\pha J}\tilde{e}^J+2i\tilde{\epsilon}\,e^I   ~. \label{Eq:GaugeTransform_Bitetrad_CommutativeLimit2}
\end{align}
Thus, a generic gauge transformation acts on the bitetrad as the composition of a Lorentz transformation and a transformation  generated by $\gamma_5$.  The former treats both the polar and the axial components of the bitetrad on the same footing, whereas the latter introduces a mixing between the two.
We note that, if we demand that the tetrads be real, then we must require that $\tilde{\epsilon}$ is pure imaginary as a consistency condition. 
In the remainder of this section, we will study the transformation properties of the metric tensors under Abelian gauge transformations generated by $\gamma_5$.

The metric tensors can be defined using the two tetrad frames as follows
\begin{align}
g_{ab}=\eta_{IJ}e^I_a e^J_b ~, \label{Eq:DefineMetric} \\
\tilde{g}_{ab}=\eta_{IJ}\tilde{e}^I_a \tilde{e}^J_b ~.\label{Eq:DefineTildeMetric}
\end{align}
The relation between the two metrics can be established using Eq.~(\ref{Eq:GeneralStarDecompositionTetrads}). In fact, in the commutative limit the assumption $M^{I}_{\pha J} \in  \widehat{ \rm GL(2,\mathbb{C})}$ amounts to the following no-shear condition
\be\label{Eq:GL(2,C)ansatz_commutative}
\tilde{e}^I_a=\Omega\, \Lambda^{I}_{\pha J}e^J_a ~
\ee
with 
$\Lambda^{I}_{\pha J}$  a local Lorentz transformation and $\Omega$ a real function representing a Weyl rescaling. Therefore, from Eq.~(\ref{Eq:GL(2,C)ansatz_commutative}) and the definitions (\ref{Eq:DefineMetric}), (\ref{Eq:DefineTildeMetric}), one has
\be\label{Eq:ConformallyRelatedMetrics}
\tilde{g}_{ab}=\Omega^2\, g_{ab}~.
\ee
Thus, the two metric tensors are conformally related and $\Omega^2$ is a positive function representing their relative scale.
The finite form of an Abelian gauge transformation generated by $\gamma_5$ acts on the bitetrad components as follows
 \begin{align}
e_a^I&\rightarrow \cos(2\tilde{\epsilon}) ~e_a^I +i\sin(2\tilde{\epsilon})~{\tilde e}_a^I   ~,\\
{\tilde e}_a^I&\rightarrow \cos(2\tilde{\epsilon}) ~ \tilde{e}_a^I +i\sin(2\tilde{\epsilon})~e_a^I   ~.
\end{align}
Upon defining $\chi=2i\tilde{\epsilon}$, and assuming that $\chi$ be real (see discussion above), we obtain
\begin{align}
e_a^I&\rightarrow \cosh\chi ~e_a^I +\sinh\chi~{\tilde e}_a^I  ~, \label{Eq:TetradChiralTransf1}\\
{\tilde e}_a^I&\rightarrow \cosh\chi ~ \tilde{e}_a^I +\sinh\chi~e_a^I   ~. \label{Eq:TetradChiralTransf2}
\end{align}
Consequently, the metric tensors transform as
\begin{align}
g_{ab}&\rightarrow \cosh^2\chi\; g_{ab}+\sinh^2\chi\;\tilde{g}_{ab}+\cosh\chi\sinh\chi\;\eta_{IJ}(\tilde{e}^I_a e^J_b+e^I_a \tilde{e}^J_b)~,\label{Eq:MetricFiniteChiralAbelianSymmetry1}  \\
\tilde{g}_{ab} &\rightarrow \cosh^2\chi\; \tilde{g}_{ab}+\sinh^2\chi\;g_{ab}+\cosh\chi\sinh\chi\;\eta_{IJ}(\tilde{e}^I_a e^J_b+e^I_a \tilde{e}^J_b) ~.\label{Eq:MetricFiniteChiralAbelianSymmetry2}
\end{align}
Using Eq.~(\ref{Eq:GL(2,C)ansatz_commutative}), and considering for simplicity an infinitesimal Lorentz transformation $\Lambda^{I}_{\pha J}\simeq\delta^{I}_{\pha J}+\epsilon^{I}_{\pha J}$,  we obtain from Eqs.~(\ref{Eq:MetricFiniteChiralAbelianSymmetry1}), (\ref{Eq:MetricFiniteChiralAbelianSymmetry2}) the following transformation laws
\begin{align}
g_{ab}&\rightarrow(\cosh\chi+\Omega\sinh\chi)^2\, g_{ab} \label{Eq:MetricRescalingFirstOrder1}\\
\tilde{g}_{ab}&\rightarrow(\cosh\chi+\Omega^{-1}\sinh\chi)^2\, \tilde{g}_{ab} \label{Eq:MetricRescalingFirstOrder2}~.
\end{align}
Accordingly, the relative scale $\Omega^2$ of the two metrics transforms as
\be\label{Eq:TransformationScaleOmega}
\Omega^2 \rightarrow \left(\frac{\Omega+\tanh\chi}{1+\Omega\tanh\chi}\right)^2~.
\ee
We observe that for $\chi\to\pm\infty$ the relative scale approaches unity, i.e.~$\Omega\to1$, which is to be expected from the form of Eqs.~(\ref{Eq:TetradChiralTransf1}), (\ref{Eq:TetradChiralTransf2}). Moreover, for $\Omega^2<1$ there is a particular value of $\chi$ such that the r.h.s.\ of (\ref{Eq:TransformationScaleOmega}) vanishes. Similarly, for $\Omega^2>1$ the denominator of the r.h.s.\ of (\ref{Eq:TransformationScaleOmega}) has a pole for some finite $\chi$.

\subsection{Action Principle}
The reformulation of the Holst action using the spinorial representation of the Lorentz group, Eq.~(\ref{Eq:AC_FullAction_Commutative}), admits a straightforward generalization to the noncommutative case
\be\label{Eq:AC_FullAction}
S[{\bm e},{\bm \omega}]=\frac{i}{32\pi G}\int \Tr\left[{\bm e}\wedge_{\star}{\bm e}\wedge_{\star}\left( *_{\Hdg}{\bm F}+\frac{1}{\beta}{\bm F}\right) \right]~
\ee
where the Hodge star operator is now \emph{defined} by Eq.~\eqn{Hstardef} through the matrix $\gamma_5$ and $\wedge_\star$ is given by Eq. \eqn{Eq:DeformedWedgeProd_Define}. 
The dynamical variables are the bitetrad ${\bm e}$, defined in Eq.~(\ref{Eq:RepeatDefinition_Bitetrad}), and the ${\mathfrak{ gl}}(2,\mathbb{C})$ gauge connection ${\bm \omega}$, defined in Eq.~(\ref{Eq:RepeatDefinition_Omega}). The field strength of the latter was defined in Eq.~(\ref{Eq:FieldStrenghtDecomposition}).
The action \eqn{Eq:AC_FullAction} is invariant under the gauge transformations \eqn{Eq:BiTetrad_GaugeTransform}--\eqn{Eq:Curvature_GaugeTransform} (see Section \ref{symm}). Moreover, it is invariant with respect to diffeomorphisms and $\star$-diffeomorphisms. The latter can be easily shown  by trivially extending the proof given in \cite{gespo2} in Appendix A.3 to the complete Palatini-Holst action. The reader is referred to Ref.~\cite{gespo2}  for details.

 As shown in the previous section, the enlargement of the internal gauge symmetry from the Lorentz symmetry to $\mathfrak{ gl}(2,\mathbb{C})$ is the minimal choice which is compatible with the twist. 
This in turn leads to the introduction of new gravitational degrees of freedom, represented by the extra components of  ${\bm e}$ and ${\bm \omega}$, which, thanks to   simple requirements of mathematical consistency have  led to a bimetric theory of gravity. 
Therefore, differently from Ref.~\cite{aschieri} we shall not impose that the extra degrees of freedom vanish in the commutative limit, but we shall retain all 
the extra fields as physical and 
provide an interpretation in the framework of the bimetric theory envisaged above. We note that the theory will naturally feature higher-order derivatives as a consequence of the twist deformation (see Appendix \ref{noncommapp}).

\section{Noncommutative gravity with a self-dual connection}\label{noncomm}
It is well known that in the commutative Palatini-Holst theory the choice $\beta=-i$ leads to a theory of a self-dual connection. This result also holds in the noncommutative theory considered. The generalization of the notion of self-duality of the gauge connection to the noncommutative case is straightforward. In fact, the definition given in the commutative case, $*_{\Hdg}{\bm \omega}\coloneqq -i{\bm \omega}\gamma_5$, is still a sound one after enlarging the gauge symmetry, with the $\mathfrak{ gl}(2,\mathbb{C})$ gauge connection ${\bm \omega}$ given by Eq.~(\ref{Eq:RepeatDefinition_Omega}). Also in this case, any connection ${\bm \omega}$ is uniquely decomposed into its self-dual and anti self-dual parts as
\be
{\bm \omega}={\bm \omega}_+ +{\bm \omega}_-~,
\ee
with ${\bm \omega}_+$, ${\bm \omega}_-$ defined as in Eq.~(\ref{Eq:LeftHanded_SelfDual}), and satisfying Eq.~(\ref{Eq:Self-Duality_EigenvalueEquation}).
Again, it can be shown that the field strength can be uniquely decomposed as
\be\label{Eq:FieldStrength_SelfPlusAntiSelf}
{\bm F}({\bm \omega})=\left(\de{\bm \omega}-i{\bm \omega}\wedge_{\star}{\bm \omega}\right)=\left(\de{\bm \omega}_+ -i{\bm \omega}_+\wedge_{\star}{\bm \omega}_+\right) +\left(\de{\bm \omega}_-  -i {\bm \omega}_{-}\wedge_{\star}{\bm \omega}_{-}\right)\equiv {\bm F}_+({\bm \omega}_+) +{\bm F}_-({\bm \omega}_-) ~,
\ee
where we defined, respectively, the self-dual and the anti self-dual field strengths
\begin{align}
{\bm F}_+({\bm \omega}_+)&=\de{\bm \omega}_+ -i{\bm \omega}_+\wedge_{\star}{\bm \omega}_+  ~,\\
{\bm F}_-({\bm \omega}_-)&=\de{\bm \omega}_-  -i{\bm \omega}_-\wedge_{\star}{\bm \omega}_-  ~.
\end{align}
Self-duality of the ${\bm F}_+$ can be proved by observing that
\be
P_+{\bm F}_+ P_+=\de(P_+{\bm \omega}_+) -i(P_+)^2{\bm \omega}_+\wedge_{\star}{\bm \omega}_+=\de{\bm \omega}_+  -i(P_+{\bm \omega}_+)\wedge_{\star}(P_+{\bm \omega}_+)={\bm F}_+~.
\ee
Similarly, it can be shown that ${\bm F}_-$ is anti self-dual.
We observe that there are no mixed terms in Eq.~(\ref{Eq:FieldStrength_SelfPlusAntiSelf}), since
\be
{\bm \omega}_+\wedge_{\star}{\bm \omega}_-=({\bm \omega}P_+)\wedge_{\star}({\bm \omega}P_-)=(P_+ P_-){\bm \omega}\wedge_{\star}{\bm \omega}=0~.
\ee
Similarly, it can be shown that ${\bm \omega}_-\wedge_{\star}{\bm \omega}_+=0$. The result expressed by Eq.~(\ref{Eq:FieldStrength_SelfPlusAntiSelf}) crucially depends on the fact that all of the $\mathfrak{ gl}(2,\mathbb{C})$ algebra generators commute with the $\gamma_5$ matrix and, hence, with the projectors $P_+$, $P_-$.

A self-dual  $\mathfrak{gl}(2,\mathbb{C})$ connection and its field strength must satisfy the following algebraic equations
\begin{align}
{\bm \omega}_+ &=-{\bm \omega}_+ \gamma_5 ~\label{Eq:SD_algebraic1} \\
{\bm F}_+ &= -{\bm F}_+ \gamma_5 ~,\label{Eq:SD_algebraic2}
\end{align}
which follow from Eq.~(\ref{Eq:Self-Duality_EigenvalueEquation}).
We remark that Eqs.~(\ref{Eq:SD_algebraic1}), (\ref{Eq:SD_algebraic2}) are clearly preserved under deformed gauge transformations (\ref{Eq:Connection_GaugeTransform}), (\ref{Eq:Curvature_GaugeTransform}).
These equations imply that a self-dual  $\mathfrak{gl}(2,\mathbb{C})$ connection has the following expansion in components
\be\label{Eq:SelfDual_Omega_Expansion}
{\bm \omega}_+=\frac{1}{2}\omega^{IJ}_+\Gamma_{IJ}+\omega_+ (\mathbb{I}-\gamma_5)~,
\ee
with $\omega^{IJ}_+$ being a self-dual spin connection, i.e.~satisfying
\be
\omega^{IJ}_+=-\frac{i}{2}\varepsilon^{IJ}_{\pha\pha KL}\,\omega^{KL}_+~.
\ee
Direct comparison between Eq.~(\ref{Eq:SelfDual_Omega_Expansion}) and the general expression (\ref{Eq:RepeatDefinition_Omega}) shows that we also have $\tilde{\omega}=-\omega$, which explains the second term in Eq.~(\ref{Eq:SelfDual_Omega_Expansion}).
Since the field strength ${\bm F}_+$ of ${\bm \omega}_+$ is also self-dual, it admits the following expansion in components
\be
{\bm F}_+=\frac{1}{2}F^{IJ}_+\Gamma_{IJ}+r_+ (\mathbb{I}-\gamma_5)~,
\ee
with
\begin{align}
F^{IJ}_+=&\de \omega_+^{IJ}+\frac{1}{2}\left((\omega_+)^I_{\pha K} \wedge_{\star} \omega_+^{KJ}-(\omega_+)^J_{\pha K} \wedge_{\star} \omega_+^{KI}\right)-i(\omega_+^{IJ}\wedge_{\star}\omega_+ +\omega_+\wedge_{\star}\omega_+^{IJ})\label{Eq:Eq:FieldStrenghtComponents1}\\
r_+=&\de\omega_+-\frac{i}{8}(\omega_+^{IJ}\wedge_{\star} (\omega_+)_{IJ})-2i(\omega_+\wedge_{\star}\omega_+)  ~.\label{Eq:Eq:FieldStrenghtComponents2}
\end{align}
The reader may appreciate the remarkable simplifications following from the self-duality condition by comparing Eqs.~(\ref{Eq:Eq:FieldStrenghtComponents1}), (\ref{Eq:Eq:FieldStrenghtComponents2}) with the components expressions of a general $\mathfrak{gl}(2,\mathbb{C})$ gauge connection, Eqs.~(\ref{Eq:Eq:FieldStrenghtDecompositionExpanded1}), (\ref{Eq:Eq:FieldStrenghtDecompositionExpanded2}), (\ref{Eq:Eq:FieldStrenghtDecompositionExpanded3}). 

Having clarified the relation between self-duality of the $\mathfrak{gl}(2,\mathbb{C})$ gauge connection and left-handedness, it is convenient to use the projectors $P_+$, $P_-$ to decompose the bitetrad (\ref{Eq:TetradDecompose}) in order to show how they couple to the field strength. Thus, we have the following decomposition
\be
{\bm e}=P_+\,{\bm u}\,P_- + P_-\,{\bm v}\,P_+~,
\ee
where we defined
\begin{align}
{\bm u}&=u^I\gamma_I~,\hspace{1em} u^I=e^I+\tilde{e}^I ~,\\
{\bm v}&=v^I\gamma_I~,\hspace{1em} v^I=e^I-\tilde{e}^I ~.
\end{align}

Going back to the action (\ref{Eq:AC_FullAction}), we observe that for $\beta=-i$ it depends only on the self-dual part of the gauge connection. Thus, we obtain
\be\label{eq:FullActionNCSelfDual}
S[{\bm e},{\bm \omega_+}]=-\frac{1}{16\pi G}\int \Tr\left[{\bm e}\wedge_{\star}{\bm e}\wedge_{\star}{\bm F}_+(\omega_+) \right]=-\frac{1}{16\pi G}\int \Tr\left[{\bm u}\wedge_{\star}{\bm v}\wedge_{\star}{\bm F}_+(\omega_+) \right] ~.
\ee
Evaluating the trace in Eq.~(\ref{eq:FullActionNCSelfDual}) we can recast the action in the form
\be\label{Eq:NonCommutativeAction}
S[{\bm e},{\bm \omega_+}]=\frac{i}{8\pi G}\int u^I\wedge_{\star} v^J\wedge_{\star}\Big[(F_+)_{IJ}+2i\eta_{IJ}r_+\Big]~.
\ee
 Similarly, for $\beta=i$ the action (\ref{Eq:AC_FullAction}) turns out to depend only on the anti self-dual part of the gauge connection, thus leading to
\be\label{Eq:NonCommutativeActionMinus}
S[{\bm e},{\bm \omega_-}]=\frac{1}{16\pi G}\int \Tr\left[{\bm e}\wedge_{\star}{\bm e}\wedge_{\star}{\bm F}_-(\omega_-) \right]=-\frac{i}{8\pi G}\int v^I\wedge_{\star} u^J\wedge_{\star}\Big[(F_-)_{IJ}+2i\eta_{IJ}r_-\Big]~.
\ee
\subsection{Equations of motion}\label{sec:eom}
Henceforth we will focus on the self-dual case. We can obtain the first set of equations of motion in this theory by varying the action (\ref{Eq:NonCommutativeAction}) w.r.t.~$u^I$ and $v^I$. Thus, we get the deformed self-dual field equations
\begin{align}
v^J \wedge_{\star}\Big[(F_+)_{IJ}+2i\eta_{IJ}\,r_+\Big]&=0~,\label{Eq:NCcaseFieldEquation1}\\
\Big[(F_+)_{IJ}-2i\eta_{IJ}\,r_+\Big]\wedge_{\star}u^J &=0~. \label{Eq:NCcaseFieldEquation2}   
\end{align}
Before deriving the next set of equations of motion we introduce the following shorthand notation for the symmetric and the antisymmetric parts (w.r.t.~the internal indices) of $u^{I}\wedge_{\star}v^{J}$
\begin{align}
K^{IJ}&=u^{(I}\wedge_{\star}v^{J)}~.\label{Eq:DefineK}\\
B^{IJ}&=u^{[I}\wedge_{\star}v^{J]}~,\label{Eq:DefineB}
\end{align}
Recalling (\ref{Eq:Eq:FieldStrenghtComponents1}), (\ref{Eq:Eq:FieldStrenghtComponents2}), we compute the variation of the action w.r.t.~$\omega_+$, which yields
\be\label{Eq:NCcaseWedgeTetradsEquation}
\de K^{I}_{\pha I}-2i\left(\omega_+\wedge_{\star}K^{I}_{\pha I}-K^{I}_{\pha I}\wedge_{\star}\omega_+\right)-\frac{1}{2}\left[(\omega_+)_{IJ}\wedge_{\star}B^{IJ}-B^{IJ}\wedge_{\star}(\omega_+)_{IJ}\right]=0~.
\ee
Similarly, variation w.r.t.~$\omega_+^{IJ}$ leads to
\be\label{Eq:NCcaseConnectionEquation}
\de B^{IJ}+\omega_+^{[J | K}\wedge_{\star}B^{I]}_{\pha K}-B_K^{\pha [J}\wedge_{\star}\omega_+^{K | I]}-i\left(\omega_+\wedge_{\star}B^{IJ}-B^{IJ}\wedge_{\star}\omega_+ \right)+\frac{1}{4}\left(\omega_+^{IJ}\wedge_{\star}K^L_{\pha L}-K^L_{\pha L}\wedge_{\star}\omega_+^{IJ}\right)=0~.
\ee
Equation (\ref{Eq:NCcaseConnectionEquation}) represents a generalization of the equation of motion for the spin connection in standard Palatini gravity, as we will discuss in more detail in Section \ref{commlim}.

\subsection{Symmetries: deformed gauge invariance and duality}\label{symm}
The noncommutative gravity theory given by the action (\ref{Eq:AC_FullAction}) exhibits several symmetries, both continuous (space-time and gauge symmetries) and discrete (dualities). We will start with a brief review of the gauge symmetries of the theory. This will be followed by a discussion of novel dualities exhibited by the model, which will be the main focus of this Section.

 Gauge symmetries of the theory are noncommutative generalizations of the symmetries of the standard Palatini theory. In particular, as shown in Section \ref{Section:NCextension}, consistency with the noncommutative product of fields requires an enlargement of the internal symmetry group from the Lorentz group to ${\rm GL}(2,\C)$. The theory also exhibits two different kinds of spacetime symmetries. In fact, the action~(\ref{Eq:AC_FullAction}) is clearly diffeomorphism invariant. However, it fails to be background independent due to the presence of the non-dynamical background fields $X_\alpha$, which appear in the definition of the twist (see Appendix~\ref{noncommapp}).
Moreover, the theory is also invariant under the action of \emph{deformed diffeomorphisms}. These are $\star$-deformations of infinitesimal diffeomorphisms, obtained by composing the Lie derivative with the twist \cite{aschieri_diffeo_twist,aschieri_stardiffeo}. 
The proof of the invariance of the action (\ref{Eq:AC_FullAction}) under infinitesimal  $\star$-diffeomorphisms is a straightforward generalization of the one given in Refs.~\cite{aschieri,gespo2} for the noncommutative Palatini action.

Infinitesimal gauge transformations of the basic dynamical variables are given by Eqs.~(\ref{Eq:BiTetrad_GaugeTransform}), (\ref{Eq:Connection_GaugeTransform}), (\ref{Eq:Curvature_GaugeTransform}). We observe that the generalization of the standard Hodge duality is compatible with deformed gauge symmetry.
In fact, the dual of the field strength transforms as
\be
\delta *_{\Hdg}{\bm F}=\delta (-i{\bm F}\gamma_5)=*_{\Hdg}\delta{\bm F} ~,
\ee
since the generators of the gauge group commute with $\gamma_5$. We introduce the shorthand notation
\be
K {\bm F}\coloneqq *_{\Hdg}{\bm F}+\frac{1}{\beta}{\bm F} ~.
\ee
Thus, we have 
\be
\delta( K {\bm F})= K\delta {\bm F} ~.
\ee
It is then staightforward to compute the variation of the action (\ref{Eq:AC_FullAction}) under a gauge transformation, which gives
\begin{align}
\delta S[{\bm e},{\bm \omega}]&=\frac{i}{32\pi G}\int \Tr\left[\delta{\bm e}\wedge_{\star}{\bm e}\wedge_{\star}K {\bm F}+{\bm e}\wedge_{\star}\delta{\bm e}\wedge_{\star}K {\bm F}+{\bm e}\wedge_{\star}{\bm e}\wedge_{\star}K \delta{\bm F}\right]=\\
&=-\frac{1}{32\pi G}\int \Tr\Big[[{\bm e},{\bm \epsilon}]_{\star}\wedge_{\star}{\bm e}\wedge_{\star}K {\bm F}+{\bm e}\wedge_{\star}[{\bm e},{\bm \epsilon}]_{\star}\wedge_{\star}K {\bm F}+{\bm e}\wedge_{\star}{\bm e}\wedge_{\star}K [{\bm F},{\bm \epsilon}]_{\star}\Big]=0 ~.
\end{align}
In the last step we used the ciclicity of the trace and the graded ciclicity property (\ref{Eq:GradedCiclicity}) of the $\wedge_{\star}$ product, along with properties (\ref{Eq:CompatibilityWedgeStar_StarProd1}), (\ref{Eq:CompatibilityWedgeStar_StarProd2}).

The theory described by the action (\ref{Eq:AC_FullAction}) exhibits three different kinds of dualities. These are all target space dualities. In the following we discuss them separately.
\begin{enumerate}[label=\roman*)]
\item The first duality is a straightforward generalization of the usual Hodge duality of the Holst theory. 

Namely, the transformation
\be
{\bm \omega}\to *_{\Hdg}{\bm \omega} ~
\ee
has the effect of exchanging the Palatini and the Holst term in the action (\ref{Eq:AC_FullAction}), and is equivalent to the following transformation of the couplings of the model
\begin{align}
G&\to \beta G\\
\beta&\to -\frac{1}{\beta} ~.\label{Eq:ImmirziTransformation}
\end{align}
The transformation (\ref{Eq:ImmirziTransformation}) has two fixed points at $\beta=\pm i$, which correspond to the (anti) self-dual theory.

\item The second type of duality corresponds to the exchange of the polar and axial components of the bitetrad
\begin{align}
{\bm e}_P\to {\bm e}_A  ~, \label{Eq:DualityTypeII_1}\\
{\bm e}_A\to {\bm e}_P  ~, \label{Eq:DualityTypeII_2}
\end{align}
which can be expressed more compactly as
\be\label{Eq:duality2}
{\bm e}\to {\bm e}\gamma_5 ~.
\ee
An alternative form for the transformation laws (\ref{Eq:DualityTypeII_1}), (\ref{Eq:DualityTypeII_2}) is
\begin{align}
{\bm u} &\to {\bm u}~,\\
{\bm v} &\to -{\bm v} ~.
\end{align}
It follows from Eq.~(\ref{Eq:duality2}) that
\be
{\bm e}\wedge_{\star} {\bm e}\to -{\bm e}\wedge_{\star} {\bm e} ~.
\ee
Hence, the action is invariant (up to a sign) and the dynamics of the pure gravity theory is clearly invariant.

\item Lastly, we consider the transformation
\begin{align}
{\bm u} &\to  \gamma_0{\bm v}\gamma_0 ~,\label{Eq:DualityTypeIII_1}\\
{\bm v} &\to \gamma_0{\bm u}\gamma_0 ~, \label{Eq:DualityTypeIII_2} \\
{\bm F} & \to \gamma_0{\bm F}\gamma_0 ~. \label{Eq:DualityTypeIII_3}
\end{align}
The matrix $\gamma_0$ implements parity in the internal space. Such a transformation has the effect of exchanging the roles of ${\bm u}$ and ${\bm v}$, while flipping the chiralities of all fields. Equations (\ref{Eq:DualityTypeIII_1}), (\ref{Eq:DualityTypeIII_2}) imply
\be
{\bm e} \to \gamma_0 {\bm e} \gamma_0 ~,
\ee
which can be expressed in components as
\begin{align}
&e^0 \to e^0  ~~~, ~~~ e^i \to -e^i ~,\\
&\tilde{e}^0 \to -\tilde{e}^0 ~, ~~~ \tilde{e}^i \to \tilde{e}^i ~.
\end{align}
Under such a transformation, the integrand in the action (\ref{Eq:AC_FullAction}) transforms as
\be
\Tr\left[{\bm e}\wedge_{\star}{\bm e}\wedge_{\star}\left( *_{\Hdg}{\bm F}+\frac{1}{\beta}{\bm F}\right) \right] \to - \Tr\left[{\bm e}\wedge_{\star}{\bm e}\wedge_{\star}\left( *_{\Hdg}{\bm F}-\frac{1}{\beta}{\bm F}\right) \right] ~,
\ee
where we used
\be
*_{\Hdg}{\bm F}  \to -i(\gamma_0{\bm F}\gamma_0)\gamma_5 =-\gamma_0 (-i{\bm F}\gamma_5)\gamma_0=   -\gamma_0(*_{\Hdg}{\bm F})\gamma_0 ~.
\ee
Hence, the transformation laws (\ref{Eq:DualityTypeIII_1}), (\ref{Eq:DualityTypeIII_2}), (\ref{Eq:DualityTypeIII_3}) determine a new duality symmetry, which leaves the equations of motion invariant while flipping the sign of the Barbero-Immirzi parameter $\beta\to - \beta$. As a particular case, the self-dual action obtained for $\beta=-i$ is dual to the anti self-dual one corresponding to $\beta=+i$.

\end{enumerate}

\section{The commutative limit}\label{commlim}
The commutative limit of the theory is formally obtained by letting the deformation parameter tend to zero $\theta^{\alpha\beta}\to0$. In this limit, Eqs.~(\ref{Eq:DefineB}), (\ref{Eq:DefineK}) become
\begin{align}
 K^{IJ}&=-2\, e^{(I}\wedge \tilde{e}^{J)}   ~,\\
 B^{IJ}&= e^I\wedge e^J-\tilde{e}^I\wedge \tilde{e}^J ~.
 \end{align}
 Thus, the equations of motion (\ref{Eq:NCcaseWedgeTetradsEquation}), (\ref{Eq:NCcaseConnectionEquation}), (\ref{Eq:NCcaseFieldEquation1}), (\ref{Eq:NCcaseFieldEquation2}) boil down to 
 \begin{align}
 & \de K^{I}_{\pha I} =0 \label{Eq:EOM1_Commutative}\\
 & \de_{\omega}B^{IJ} \coloneqq \de B^{IJ}+\omega_+^{[J | K}\wedge B^{I]}_{\pha K}-B_K^{\pha [J}\wedge\omega_+^{K | I]} =0 \label{Eq:Connection_CommutativeCase} \\
 & u^J \wedge\Big[(F_+)_{IJ}+2i\eta_{IJ}\,r_+\Big]=0 \label{Eq:EinsteinU_CommutativeCase} \\
 & v^J \wedge\Big[(F_+)_{IJ}-2i\eta_{IJ}\,r_+\Big]=0 \label{Eq:EinsteinV_CommutativeCase}  ~.
 \end{align}
 Let us examine these equations in detail in the following subsections.
 
 \subsection{Bitetrad constraint}
The first equation of motion, Eq.~(\ref{Eq:EOM1_Commutative}), represents a dynamical constraint on the two tetrads, which reads as
\begin{align}
\de\left(e_I\wedge \tilde{e}^I\right)=0\label{Eq:WedgeTetradsClosed} ~.
\end{align}
Equation (\ref{Eq:WedgeTetradsClosed}) means that the two-form $e_I\wedge \tilde{e}^I$ is closed; hence, it is locally exact. Therefore, we have (at least locally)
\be\label{Eq:WedgeTetradsExact}
e_I\wedge \tilde{e}^I=\de p ~.
\ee
Writing spacetime indices explicitly, Eq.~(\ref{Eq:WedgeTetradsExact}) reads as
\be
\eta_{IJ}e^I_{[a}\tilde{e}^J_{b]}=\pa_{[a}p_{b]} ~.
\ee
The two-form $\de p$ can be decomposed using the anholonomic basis obtained from the tetrad $e^I$
\be\label{Eq:DiffP_TetradBasis}
\de p=\lambda_{IJ}\,e^I\wedge e^J~, 
\ee
where the components $\lambda_{IJ}$ are given by
\be\label{Eq:Lambda_Coeff}
\lambda_{IJ}=e^{a}_{[I} e^{b}_{J]} \pa_{[a}p_{b]} ~.
\ee

We recall that the second tetrad $\tilde{e}^I$ is related to the first tetrad via the no-shear condition (\ref{Eq:GL(2,C)ansatz_commutative})
\be\label{Eq:ProjectTildeTetrad}
\tilde{e}^I=M^{I}_{\pha J} e^J= \Omega \Lambda^{I}_{\pha J} e^J ~,
\ee
where $\Omega$ and $\Lambda^{I}_{\pha J}$ are both spacetime dependent.
Plugging Eq.~(\ref{Eq:ProjectTildeTetrad}) in Eq.~(\ref{Eq:WedgeTetradsExact}), we obtain
\be\label{Eq:AntisymmetricPartM_Exact}
\de p=M_{[IJ]}e^I\wedge e^{J} ~.
\ee
Therefore, comparing Eq.~(\ref{Eq:DiffP_TetradBasis}) and Eq.~(\ref{Eq:AntisymmetricPartM_Exact}), and using Eq.~(\ref{Eq:Lambda_Coeff}), we conclude
\be\label{Eq:RelationM_lambda}
M_{[IJ]}=\lambda_{IJ}=e^{a}_{[I} e^{b}_{J]} \pa_{[a}p_{b]}~.
\ee
This result shows that the antisymmetric part of the matrix $M_{IJ}$ in Eq.~(\ref{Eq:ProjectTildeTetrad}) is determined once a one-form $p$ is assigned.

\subsection{Connection Equation}\label{Sec:ConnectionEq}

From the equation of motion~(\ref{Eq:Connection_CommutativeCase}), adopting the definitions $T^I \coloneqq \de_\omega e^I$, $\tilde{T}^I \coloneqq \de_\omega \tilde{e}^I$, we obtain
\be\label{EQ:eomCommutative}
e^{[I}\wedge T^{J]}-\tilde{e}^{[I}\wedge \tilde{T}^{J]}=0.
\ee
Equation (\ref{EQ:eomCommutative}) is a generalization of the equation of motion for the spin connection in standard Palatini theory.
Using the no-shear condition (\ref{Eq:ProjectTildeTetrad}), equation (\ref{EQ:eomCommutative}) implies
\be\label{Eq:EquationTorsionwithDilation}
\left(\delta^{IJ}_{\pha\pha HK}-\Omega^2 \Lambda^I_{\pha [H}\Lambda^J_{\pha K]}\right) e^H\wedge T^K+\Omega^2 \Lambda^I_{\pha [H}\Lambda^J_{\pha K]}\de( \log \Omega)\wedge e^H \wedge  e^K=0~.
\ee
It is convenient to define
\be\label{Eq:DefineQoperator}
Q^{IJ}_{\pha\pha HK}=\delta^{IJ}_{\pha\pha HK}-\Omega^2 \Lambda^{[I}_{\pha H}\Lambda^{J]}_{\pha K} ~.
\ee
Thus, Eq.~(\ref{Eq:EquationTorsionwithDilation}) can be recast in the following form
\be\label{Eq:EquationTorsionwithDilation_Final}
Q^{IJ}_{\pha\pha HK}\left(e^H \wedge T^K -\de ( \log \Omega)\wedge e^H \wedge  e^K \right)+\de ( \log \Omega)\wedge e^I \wedge  e^J =0 ~.
\ee
The solution of Eq.~(\ref{Eq:EquationTorsionwithDilation_Final}) requires a detailed analysis. To begin with, we investigate the relation between a non-vanishing torsion $T^I$ and the dilation factor $\Omega$.
\begin{enumerate}[label=\roman*)]
\item Firstly, we assume $T^I=0$ identically in a spacetime region $\mathcal{U}$. Then, Eq.~(\ref{Eq:EquationTorsionwithDilation_Final}) simplifies to
\be
\Omega^2 \Lambda^{[I}_{\pha H}\Lambda^{J]}_{\pha K} \de ( \log \Omega)\wedge e^H \wedge  e^K=0 ~,
\ee
which in turn implies
\be\label{Eq:TorsionwithDilation_FirstCase}
\de ( \log \Omega)\wedge e^I \wedge  e^J=0 ~.
\ee
Since the tetrad $e^I$ is assumed to be non-degenerate, we conclude $\de ( \log \Omega)=0$ in $\mathcal{U}$. Therefore, $T^I=0$ implies that $\Omega$ is constant.
\item We assume that $\Omega$ be a constant in a spacetime region $\mathcal{U}$. In this case, Eq.~(\ref{Eq:EquationTorsionwithDilation_Final}) leads to
\be\label{Eq:EquationTorsionwithDilation_UnitOmega}
Q^{IJ}_{\pha\pha HK} e^H \wedge T^K =0 ~.
\ee
\begin{enumerate}[label=\alph*)]
\item If $\Omega\neq1$, the operator $Q^{IJ}_{\pha\pha HK}$ does not admit zero modes. Therefore, the only solution in this case is $T^I=0$ identically in $\mathcal{U}$.

\item The case $\Omega=1$ identically requires more care. In fact, in this case the operator $Q^{IJ}_{\pha\pha HK}$ may admit zero modes, depending on the particular Lorentz transformation $\Lambda^{I}_{\pha J}$ (assumed to be non-trivial) entering the definition (\ref{Eq:DefineQoperator}). The problem is thus reduced to finding invariant bivectors (i.e.~skew-symmetric tensors) under the Lorentz transformation $\Lambda^I_{\pha J}$ at a spacetime point $x$. A point $x$ where such invariant bivectors exist will be referred to as a \emph{critical torsion point}.
We denote by $A^{IJ}$ an invariant bivector\footnote{Given a (non-trivial) Lorentz transformation, the space of invariant bivectors is at most one-dimensional. Examples are given by: a rotation in the (1,2) plane, which leaves the plane (0,3) invariant; a boost in the 1 direction, which leaves the plane (2,3) invariant. The corresponding bivectors are skew-symmetric matrices with the only non-zero entries in correspondence with the invariant planes. We stress that a generic Lorentz transformation \emph{does not} admit invariant bivectors.} in the internal space; i.e.~a solution to the equation
\be
\Lambda^{[I}_{\pha H}\Lambda^{J]}_{\pha K} A^{HK} =A^{IJ} ~.
\ee
Let $\sigma$ be an arbitrary three-form, which we may expand in the tetrad basis as
\be
\sigma=\sigma_{IJK}e^I\wedge e^J \wedge e^K ~.
\ee
Thus, the solution of Eq.~(\ref{Eq:EquationTorsionwithDilation_UnitOmega}) in this case is given by
\be\label{Eq:EquationTorsionwithDilation_SingularSolution}
e^{[I} \wedge T^{J]}=A^{IJ} \sigma ~.
\ee
\end{enumerate}
\end{enumerate}

Going back to Eq.~(\ref{Eq:EquationTorsionwithDilation_Final}), we observe that a simple solution can be obtained for a trivial Lorentz transformation $\Lambda^I_{\pha J}(x)=\delta^I_{\pha J}$. With this assumption, Eq.~(\ref{Eq:EquationTorsionwithDilation_Final}) reduces to
\be
(1-\Omega^2) e^{[I} \wedge T^{J]}+\Omega^2 \de ( \log \Omega)\wedge e^{[I} \wedge  e^{J]}=0 ~. 
\ee
Non-degeneracy of the tetrad implies, assuming $\Omega\neq 1$:
\be\label{Eq:TorsionwithDilation_SimpleSolution}
 T^{J}=(1-\Omega^2)^{-1} \Omega^2 \, \de ( \log \Omega) \wedge  e^{J} ~.
\ee
Note that, in this particular case, vanishing torsion implies constant $\Omega$, and vice versa.

Next, we seek more general solutions of Eq.~(\ref{Eq:EquationTorsionwithDilation_Final}) featuring both a non-constant $\Omega$ and non-vanishing torsion, assuming a non-trivial Lorentz transformation $\Lambda^I_{\pha J}$. We consider a first order expansion of $\Lambda^I_{\pha J}$ around the identity
\be\label{Eq:InfinitesimalLorentz}
\Lambda^I_{\pha J}\simeq\delta^{I}_{\pha J}+\epsilon^{I}_{\pha J} ~.
\ee
Plugging this expansion in Eq.~(\ref{Eq:EquationTorsionwithDilation_Final}) we obtain
\be
\begin{split}
(1-\Omega^2) e^{[I} \wedge T^{J]}-\Omega^2\left(e^{[I}\wedge \epsilon^{J]}_{\pha H}T^H+e^{H}\epsilon^{[I}_{\pha H}\wedge T^{J]}\right)  +\\
\Omega^2 \de ( \log \Omega)\wedge \left(e^{[I} \wedge  e^{J]} +e^{[I}\wedge \epsilon^{J]}_{\pha H}e^H+e^{H}\epsilon^{[I}_{\pha H}\wedge e^{J]}\right)=0  ~.
\end{split}
\ee
Adopting a perturbative scheme, we expand the torsion around the solution corresponding to the $\epsilon^{I}_{\pha J}=0$ case
\be
T^I=T^I_{(0)}+T^I_{(1)}+\dots  ~,
\ee
where it is assumed $T^I_{(0)}=\mathcal{O}(\epsilon^0)$ and $T^I_{(1)}=\mathcal{O}(\epsilon)$. To zero-th order in perturbation theory, we have
\be\label{Eq:EquationTorsion_PerturbativeSolution_0}
(1-\Omega^2) e^{[I} \wedge T^{J]}_{(0)}+\Omega^2 \de ( \log \Omega)\wedge e^{[I} \wedge  e^{J]}=0 ~,
\ee
whose solution is given by Eq.~(\ref{Eq:TorsionwithDilation_SimpleSolution}). To first order we have
\be\label{Eq:EquationTorsion_PerturbativeSolution_1}
(1-\Omega^2) e^{[I} \wedge T^{J]}_{(1)}-\Omega^2\left(e^{[I}\wedge \epsilon^{J]}_{\pha H}T^H_{(0)}+e^{H}\epsilon^{[I}_{\pha H}\wedge T^{J]}_{(0)}\right)  +\Omega^2 \de ( \log \Omega)\wedge \left(e^{[I}\wedge \epsilon^{J]}_{\pha H}e^H+e^{H}\epsilon^{[I}_{\pha H}\wedge e^{J]}\right)=0  ~.
\ee
From Eqs.~(\ref{Eq:EquationTorsion_PerturbativeSolution_0}) and (\ref{Eq:EquationTorsion_PerturbativeSolution_1}) we obtain
\be
(1-\Omega^2) e^{[I} \wedge T^{J]}_{(1)}=e^{[I}\wedge \epsilon^{J]}_{\pha H}T^H_{(0)}+e^{H}\epsilon^{[I}_{\pha H}\wedge T^{J]}_{(0)} ~.
\ee
Thus, the effect of a Lorentz transformation in the general relation (\ref{Eq:ProjectTildeTetrad}) between the two tetrads is to give an extra contribution to the torsion. In the solution scheme adopted here,  
to zero-th order the torsion is determined by the relative scale $\Omega$ of the two tetrads and its spacetime variations, 
whereas the first order correction 
depends on the relative orientation of the two tetrads,
given by the infinitesimal Lorentz transformation in (\ref{Eq:InfinitesimalLorentz}).

\subsection{Gravitational Field Equations}
The remaining two equations of motion (\ref{Eq:EinsteinU_CommutativeCase}), (\ref{Eq:EinsteinV_CommutativeCase}) can be conveniently recast in the following form
\begin{align}
& e^J \wedge(F_+)_{IJ}-2i\,\tilde{e}_I \wedge r_+=0 \label{Eq:EinsteinU_CommutativeCase_R1} \\
 & \tilde{e}^J \wedge(F_+)_{IJ}-2i\,e_I \wedge r_+=0 \label{Eq:EinsteinV_CommutativeCase_R1}  ~.
 \end{align}
Wedge multiplying of (\ref{Eq:EinsteinU_CommutativeCase_R1}) by the tetrad $e^I$, and of Eq.~(\ref{Eq:EinsteinV_CommutativeCase_R1}) by the tilde tetrad $\tilde{e}^I$, gives
\begin{align}
& e^I\wedge e^J \wedge(F_+)_{IJ}-2i\,\de p \wedge r_+=0 \label{Eq:EinsteinU_CommutativeCase_R2} \\
 & \tilde{e}^I \wedge \tilde{e}^J \wedge(F_+)_{IJ}+2i\,\de p\wedge r_+=0 \label{Eq:EinsteinV_CommutativeCase_R2}  ~.
 \end{align}
Equations (\ref{Eq:EinsteinU_CommutativeCase_R1}), (\ref{Eq:EinsteinU_CommutativeCase_R2}) also imply
\be\label{Eq:BitetradBivectorWedgeF}
\tilde{e}^I\wedge e^J \wedge(F_+)_{IJ}=0 ~.
\ee
Using the definition of the Nieh-Yan invariant, we get from Eqs.~(\ref{Eq:EinsteinU_CommutativeCase_R2}), (\ref{Eq:EinsteinV_CommutativeCase_R2})
\begin{align}
& \de (e_I\wedge T^I)    +2i\,\de p \wedge r_+=T_I\wedge T^I \label{Eq:EinsteinU_CommutativeCase_R3} \\
 & \de (\tilde{e}_I\wedge \tilde{T}^I)    -2i\,\de p \wedge r_+=\tilde{T}_I\wedge \tilde{T}^I \label{Eq:EinsteinV_CommutativeCase_R3}  ~,
 \end{align}
 where $\tilde{T}^I\coloneqq \de_{\omega} \tilde{e}^I$. Using the 
 no-shear condition (\ref{Eq:ProjectTildeTetrad}) and Eq.~(\ref{Eq:EinsteinV_CommutativeCase_R3}), we have 
 \be\label{Eq:EinsteinV_CommutativeCase_R3_Simplified}
 \de (e_I\wedge T^I)    -2i\Omega^{-2}\de p \wedge r_+=T_I\wedge T^I  ~.
 \ee
 Thus, comparing Eqs.~(\ref{Eq:EinsteinU_CommutativeCase_R3}) and (\ref{Eq:EinsteinV_CommutativeCase_R3_Simplified}) we deduce
 \begin{align}
 &\de p \wedge r_+=0 \label{Eq:r+_Solution}\\
 &\de (e_I\wedge T^I)=T_I\wedge T^I \label{Eq:TorsionSquared_Exact}~.
 \end{align}
 Equation (\ref{Eq:TorsionSquared_Exact}) can be conveniently recast in the form
 \be
 e_I\wedge \de_{\omega}T^I =0 ~.
 \ee
 Thus, going back to Eqs.~(\ref{Eq:EinsteinU_CommutativeCase_R2}), (\ref{Eq:EinsteinV_CommutativeCase_R2}) and using Eq.~(\ref{Eq:r+_Solution}), we obtain that the the Holst densities are identically vanishing for both tetrads
\begin{align}
& e^I\wedge e^J \wedge(F_+)_{IJ}=0 \label{Eq:EinsteinU_CommutativeCase_R4} \\
 & \tilde{e}^I \wedge \tilde{e}^J \wedge(F_+)_{IJ}=0 \label{Eq:EinsteinV_CommutativeCase_R4}  ~.
 \end{align}
Using the self-duality of the field strength $F_{+}^{IJ}$, we can recast Eq.~(\ref{Eq:EinsteinU_CommutativeCase_R1}) in the equivalent form 
\be\label{Eq:EinsteinEquations_DiffForms}
\frac{1}{2}F_{+}^{IJ}\wedge e^K \eps_{IJKL}+2\tilde{e}_L \wedge r_+ =0 ~.
\ee
Writing 
spacetime indices explicitly
and after some algebraic manipulations (cf.~e.g.,~\cite{gasperini}), we are led to the following form of the field equations\footnote{Note that the tetrad is assumed  $e^{I}_{a}$ to be invertible in the derivation of Eq.~(\ref{Eq:FieldEquations_FinalForm}).}
\be\label{Eq:FieldEquations_FinalForm}
G^{ab}_+ + 2e^{a}_I \tilde{e}^{I}_{c}(* r_+)^{cb}=0 ~,
\ee
where the Hodge dual $*$ of $r_+$ is defined with respect to its \emph {spacetime indices} as
\be
(* r_+)_{ab}\coloneqq \frac{1}{2}\eps_{ab}^{\pha\pha\, cd}(r_+)_{cd} ~.
\ee
$G_{+}^{ab}$ denotes the (contravariant) Einstein tensor in a spacetime endowed with torsion (see Appendix~\ref{Appendix:Torsion}) and metric tensor given by $g_{ab}=\eta_{IJ}e^I_a e^J_b$.
Similar equations as (\ref{Eq:EinsteinEquations_DiffForms}), (\ref{Eq:FieldEquations_FinalForm}) can be written down for the dual geometry given by $\tilde{e}^{I}_{a}$ starting from Eq.~(\ref{Eq:EinsteinV_CommutativeCase_R2}).

\subsection{Reality Conditions}\label{Sec:Reality}
We now impose reality conditions, i.e.~we shall assume that the two tetrads $e^I$ and $\tilde{e}^I$ and the Weyl vector $w=-4i\,\omega_+$ be real (see Appendix~\ref{Appendix:Weyl}), and that the spin connection satisfies $\omega^{IJ}=2\,\Re\{\omega_+^{IJ}\}$ (see Section \ref{holst}). 

Thus, the tensorial form of the gravitational field equations is obtained from Eq.~(\ref{Eq:EinsteinU_CommutativeCase_R1}), by splitting it into its real and imaginary parts. 
Respectively, they read as
 \begin{align}
  &e^J \wedge F_{IJ}+\tilde{e}_I \wedge \de w=0 ~, \label{Eq:Constraint_Bigravity}\\
&\frac{1}{2}F^{IJ}\wedge e^K \eps_{IJKL}=0 ~. \label{Eq:EinsteinEquations_Forms}
\end{align}
Using the first Bianchi identity
\be\label{Eq:BianchiIdentity}
\de_\omega T^I =F^{I}_{\pha J} \wedge e^J ~,
\ee
we can recast Eq.~(\ref{Eq:Constraint_Bigravity}) in the following form
\be
\de_\omega T^I+\tilde{e}^I \wedge \de w=0 ~.
\ee
Similarly, from the real part of Eq.~(\ref{Eq:EinsteinV_CommutativeCase_R1}) we obtain
\be
\de_\omega \tilde{T}^I+e^I \wedge \de w=0 ~.
\ee
Equation (\ref{Eq:Constraint_Bigravity}) can be recast in tensor form as
\be\label{Eq:AntisymR_w}
R_{[abc]}^{\pha\pha\pha\pha d}-(\de w)_{[ab}\tilde{e}^I_{c]}e_I^d =0 ~.
\ee
Using the Bianchi identity (\ref{Eq:Bianchi1}), we can rewrite Eq.~(\ref{Eq:AntisymR_w}) as
\be\label{Eq:Relation_TorsionNonMetricity}
\nabla_{[a}T^d_{\pha bc]}-T^e_{\pha [ab}T^d_{\pha c]e}=(\de w)_{[ab}\tilde{e}^I_{c]}e_I^d ~,
\ee
where $\nabla_a$ indicates a metric compatible torsionful affine connection.
Equation (\ref{Eq:EinsteinEquations_Forms}) gives the gravitational field equation
\be\label{Eq:EinsteinCartan}
G_{ab}=0 ~,
\ee
where the Einstein tensor $G_{ab}$ includes torsion contributions, see Appendix \ref{Appendix:Torsion}.

To summarize, the set of coupled equations (\ref{Eq:WedgeTetradsClosed}), (\ref{EQ:eomCommutative}), (\ref{Eq:Relation_TorsionNonMetricity}), (\ref{Eq:EinsteinCartan}) describes the dynamics of $e^I_a$, $\tilde{e}^I_a$, $T^{a}_{\pha bc}$ and $w_a$. Solutions to the equations of motion (\ref{Eq:WedgeTetradsClosed}), (\ref{EQ:eomCommutative}) have been explictly obtained above. We note that Eq.~(\ref{Eq:EinsteinCartan}) has the same form as the field equation in vacuum Einstein-Cartan's theory. Equation~(\ref{Eq:Relation_TorsionNonMetricity}) shows that torsion is sourced by non-metricity, their interaction being mediated by the tensor $\tilde{e}^I_{a}e_I^b$. Thus, even in vacuo, torsion would be dynamical in general. This result signifies an important departure from Einstein-Cartan's theory, and is essentially due to the bimetric nature of the theory and to the presence of non-metricity.

\section{Perturbative expansion in $\theta$}\label{thetaexp}
In the previous section, we solved the equations of motion (\ref{Eq:NCcaseWedgeTetradsEquation}), (\ref{Eq:NCcaseConnectionEquation}) in the commutative case, obtained for $\theta=0$. Such equations determine the relation between the two tetrads, and the spin connection, respectively. In particular, we found that the latter admits solutions which entail significant departures from standard pure Palatini gravity. Such departures ultimately stem from the extra gravitational degrees of freedom. Our aim in this section will be to determine further corrections introduced by the noncommutative deformation. This will be done by means of a perturbative expansion in the deformation parameter~$\theta$, which is a valid approximation at scales much larger than the noncommutativity scale.

We start by expanding the two-form $u^I \wedge_\star v^J$ to second order in $\theta$ using the asymptotic expansion of the twisted wedge product $\wedge_\star$, see Eq.~(\ref{Eq:WedgeStar_Expansion})
\be
u^I \wedge_\star v^J=u^I \wedge v^J +\frac{i}{2}\,\theta^{\alpha\beta}\mathcal{L}_{X_\alpha}u^I\wedge\mathcal{L}_{X_\beta}v^J-\frac{1}{4}\,\theta^{\alpha\beta}\theta^{\gamma\delta}\mathcal{L}_{X_\alpha}\mathcal{L}_{X_\gamma}u^I\wedge\mathcal{L}_{X_\beta}\mathcal{L}_{X_\delta}v^J+\mathcal{O}(\theta^3)~.
\ee
Thus, its symmetric and antisymmetric and parts read as
 \begin{align}
 K^{IJ}&=-2e^{(I}\wedge\tilde{e}^{J)}+\frac{i}{2}\theta^{\alpha\beta}\left(\mathcal{L}_{X_\alpha}e^{I}\wedge\mathcal{L}_{X_\beta}e^{J}-\mathcal{L}_{X_\alpha}\tilde{e}^{I}\wedge\mathcal{L}_{X_\beta}\tilde{e}^{J}\right)
 +\frac{1}{2}\theta^{\alpha\beta}\theta^{\gamma\delta}\mathcal{L}_{X_\alpha}\mathcal{L}_{X_\gamma}e^{(I}\wedge\mathcal{L}_{X_\beta}\mathcal{L}_{X_\delta}\tilde{e}^{J)}+\mathcal{O}(\theta^3)~, \label{Eq:ExpansionK} \\
 B^{IJ}&= e^I\wedge e^J-\tilde{e}^I\wedge \tilde{e}^J-\frac{1}{4}\theta^{\alpha\beta}\theta^{\gamma\delta}\Big(\mathcal{L}_{X_\alpha}\mathcal{L}_{X_\gamma}e^I\wedge\mathcal{L}_{X_\beta}\mathcal{L}_{X_\delta}e^J-\mathcal{L}_{X_\alpha}\mathcal{L}_{X_\gamma}\tilde{e}^I\wedge\mathcal{L}_{X_\beta}\mathcal{L}_{X_\delta}\tilde{e}^J\Big)+\mathcal{O}(\theta^3)~. \label{Eq:ExpansionB}
  \end{align}
Note that only the trace over Lorentz indices of the two-form $K^{IJ}$ appears in the equation of motion (\ref{Eq:NCcaseWedgeTetradsEquation}). Its expression is
 \be\label{Eq:TraceKsecondOrder}
 K^I_{\pha I}=-2e^{I}\wedge\tilde{e}_{I}+\frac{i}{2}\theta^{\alpha\beta}\left(\mathcal{L}_{X_\alpha}e^{I}\wedge\mathcal{L}_{X_\beta}e_{I}-\mathcal{L}_{X_\alpha}\tilde{e}^{I}\wedge\mathcal{L}_{X_\beta}\tilde{e}_{I}\right)+\frac{1}{2}\theta^{\alpha\beta}\theta^{\gamma\delta}\mathcal{L}_{X_\alpha}\mathcal{L}_{X_\gamma}e^{I}\wedge\mathcal{L}_{X_\beta}\mathcal{L}_{X_\delta}\tilde{e}_{I}+\mathcal{O}(\theta^3)~.
 \ee
 Let us proceed by evaluating all the terms in Eq.~(\ref{Eq:NCcaseWedgeTetradsEquation})  separately. We have for the term in round brackets
 \be
 \begin{split}
 \omega_+\wedge_{\star}K^{I}_{\pha I}-K^{I}_{\pha I}\wedge_{\star}\omega_+=&\\
 & -2i\theta^{\alpha\beta}\Big(\mathcal{L}_{X_\alpha}\omega_+\wedge \mathcal{L}_{X_\beta}(e^I\wedge \tilde{e}_I)\Big)+\\
 & -\frac{1}{2}\theta^{\alpha\beta}\theta^{\rho\sigma}\mathcal{L}_{X_\beta}\left[\mathcal{L}_{X_\alpha}\omega_+ \wedge \left(\mathcal{L}_{X_\rho}e^I\wedge \mathcal{L}_{X_\sigma} e_I -\mathcal{L}_{X_\rho}\tilde{e}^I\wedge \mathcal{L}_{X_\sigma} \tilde{e}_I \right)\right]+\mathcal{O}(\theta^3)~.
 \end{split}
 \ee
 The expansion of the last two terms in Eq.~(\ref{Eq:NCcaseWedgeTetradsEquation})  gives
\be\label{Eq:OmegaBcommuteExpand}
 (\omega_+)_{IJ}\wedge_{\star}B^{IJ}-B^{IJ}\wedge_{\star}(\omega_+)_{IJ}=i\theta^{\alpha\beta}\Big(\mathcal{L}_{X_\alpha}\omega_+^{IJ}\wedge \mathcal{L}_{X_\beta}\bar{B}_{IJ}\Big)+\mathcal{O}(\theta^3)~,
 \ee
 where we defined
\be
\bar{B}^{IJ}=e^I\wedge e^J-\tilde{e}^I\wedge \tilde{e}^J~.
\ee
 Hence, up to third order terms in $\theta$, Eq.~(\ref{Eq:NCcaseWedgeTetradsEquation}) reads as
 \be\label{Eq:WedgeTetradsEquationSecondOrderThetaExpand}
  \begin{split}
\de K^{I}_{\pha I}-4\theta^{\alpha\beta}\Big(\mathcal{L}_{X_\alpha}\omega_+\wedge \mathcal{L}_{X_\beta}(e^I\wedge \tilde{e}_I)\Big)+i\theta^{\alpha\beta}\theta^{\rho\sigma}\mathcal{L}_{X_\beta}\left[\mathcal{L}_{X_\alpha}\omega_+ \wedge \left(\mathcal{L}_{X_\rho}e^I\wedge \mathcal{L}_{X_\sigma} e_I -\mathcal{L}_{X_\rho}\tilde{e}^I\wedge \mathcal{L}_{X_\sigma} \tilde{e}_I \right)\right]+ \nonumber\\
-\frac{i}{2}\theta^{\alpha\beta}\Big(\mathcal{L}_{X_\alpha}\omega_+^{IJ}\wedge \mathcal{L}_{X_\beta}\bar{B}_{IJ}\Big)=0~,
\end{split}
\ee
with $K^{I}_{\pha I}$ given by Eq.~(\ref{Eq:TraceKsecondOrder}).
 
We shall proceed similarly for Eq.~(\ref{Eq:NCcaseConnectionEquation}). For the terms in the first round bracket, we obtain
 \be
  \omega_+\wedge_{\star}B^{IJ}-B^{IJ}\wedge_{\star}\omega_+=i\theta^{\alpha\beta}\Big(\mathcal{L}_{X_\alpha}\omega_+\wedge \mathcal{L}_{X_\beta}\bar{B}^{IJ}\Big)+\mathcal{O}(\theta^3)~.
 \ee
 The second round bracket in Eq.~(\ref{Eq:NCcaseConnectionEquation}) gives
 \be
 \begin{split}
 \omega_+^{IJ}\wedge_{\star}K^L_{\pha L}-K^L_{\pha L}\wedge_{\star}\omega_+^{IJ}=&\\
 &-2i\theta^{\alpha\beta}\Big(\mathcal{L}_{X_\alpha}\omega_+^{IJ}\wedge \mathcal{L}_{X_\beta}(e^L\wedge \tilde{e}_L)\Big)+\\
& -\frac{1}{2}\theta^{\alpha\beta}\theta^{\rho\sigma}\mathcal{L}_{X_\beta}\left[\mathcal{L}_{X_\alpha}\omega_+^{IJ} \wedge \left(\mathcal{L}_{X_\rho}e^L\wedge \mathcal{L}_{X_\sigma} e_L -\mathcal{L}_{X_\rho}\tilde{e}^L\wedge \mathcal{L}_{X_\sigma} \tilde{e}_L \right)\right]+\mathcal{O}(\theta^3)~.
 \end{split}
 \ee
 The expansion of the second and third term in Eq.~(\ref{Eq:NCcaseConnectionEquation}) reads as
 \be
 \begin{split}
 \omega_+^{[J|K}\wedge_{\star}B^{I]}_{\pha K}-B_K^{\pha [J}\wedge_{\star}\omega_+^{K|I]}&=\\
&\phantom{=}
\omega_+^{[J|K}\wedge \bar{B}^{I]}_{\pha K}-\bar{B}_K^{\pha [J}\wedge\omega_+^{K|I]}+\omega_+^{[J|K}\wedge \breve{B}^{I]}_{\pha K}-\breve{B}_K^{\pha [J}\wedge\omega_+^{K|I]}+\\
&\phantom{=}
\frac{1}{2}\theta^{\alpha\beta}\theta^{\gamma\delta}\Big(\mathcal{L}_{X_\alpha}\mathcal{L}_{X_\gamma}(\omega_+)^{[I}_{\pha K}\wedge \mathcal{L}_{X_\beta}\mathcal{L}_{X_\delta}\bar{B}^{J]K}\Big)+\mathcal{O}(\theta^3)~, \label{Eq:ConnectionEquation_2ndOrderExpand_IntermediateStep}
 \end{split}
 \ee
 where
 \be
 \breve{B}^{IJ}=-\frac{1}{4}\theta^{\alpha\beta}\theta^{\gamma\delta}\Big(\mathcal{L}_{X_\alpha}\mathcal{L}_{X_\gamma}e^I\wedge\mathcal{L}_{X_\beta}\mathcal{L}_{X_\delta}e^J-\mathcal{L}_{X_\alpha}\mathcal{L}_{X_\gamma}\tilde{e}^I\wedge\mathcal{L}_{X_\beta}\mathcal{L}_{X_\delta}\tilde{e}^J\Big)~.
 \ee
Thus, Eq.~(\ref{Eq:NCcaseConnectionEquation}) reads as
\be\label{Eq:NCcaseConnectionEquationSecondOrderThetaExpand}
\begin{split}
&0=\de \left(\bar{B}^{IJ}+ \breve{B}^{IJ}\right)+\omega_+^{[J|K}\wedge \bar{B}^{I]}_{\pha K}-\bar{B}_K^{\pha [J}\wedge\omega_+^{K|I]}+\omega_+^{[J|K}\wedge \breve{B}^{I]}_{\pha K}-\breve{B}_K^{\pha [J}\wedge\omega_+^{K|I]}+\\
&
\frac{1}{2}\theta^{\alpha\beta}\theta^{\gamma\delta}\Big(\mathcal{L}_{X_\alpha}\mathcal{L}_{X_\gamma}(\omega_+)^{[I}_{\pha K}\wedge \mathcal{L}_{X_\beta}\mathcal{L}_{X_\delta}\bar{B}^{J]K}\Big)+\theta^{\alpha\beta}\Big(\mathcal{L}_{X_\alpha}\omega_+\wedge \mathcal{L}_{X_\beta}\bar{B}^{IJ}\Big)+\\
&-\frac{i}{2}\theta^{\alpha\beta}\Big(\mathcal{L}_{X_\alpha}\omega_+^{IJ}\wedge \mathcal{L}_{X_\beta}(e^L\wedge \tilde{e}_L)\Big) -\frac{1}{8}\theta^{\alpha\beta}\theta^{\rho\sigma}\mathcal{L}_{X_\beta}\left[\mathcal{L}_{X_\alpha}\omega_+^{IJ} \wedge \left(\mathcal{L}_{X_\rho}e^L\wedge \mathcal{L}_{X_\sigma} e_L -\mathcal{L}_{X_\rho}\tilde{e}^L\wedge \mathcal{L}_{X_\sigma} \tilde{e}_L \right)\right]\\
\end{split}
\ee
Equation (\ref{Eq:NCcaseConnectionEquationSecondOrderThetaExpand}) is clearly not covariant under internal Lorentz transformations, due to the $\theta$-dependent correction terms\footnote{More precisely, this is due to the appearance of the Lie derivatives, which do not transforms covariantly under internal gauge transformations.}. In fact, in the noncommutative case, $\omega_+^{IJ}$ is only a component of the $\frak{gl}(2,\C)$ gauge connection ${\bm \omega}_+$; as such, it gets mixed with the component $\omega_+$ under gauge transformations (see Eqs.~(\ref{Eq:SelfDual_Omega_Expansion}), (\ref{Eq:Connection_GaugeTransform})). From a phenomenological point of view, this is interpreted as a violation of Lorentz symmetry in the dynamics of the spin connection.

\subsection{Perturbative solutions}\label{pertsol}
We are now in a position to solve (\ref{Eq:WedgeTetradsEquationSecondOrderThetaExpand}), (\ref{Eq:NCcaseConnectionEquationSecondOrderThetaExpand}) perturbatively. We will determine perturbative corrections to solutions of the $\theta=0$ case, up to second order terms in $\theta$.
 We start by expanding the second tetrad $ \tilde{e}^I$ and torsion $T^I$ around a solution of the commutative case
 \begin{align}
  \tilde{e}^I=\tilde{e}^I_{(0)}+ \tilde{e}^I_{(1)}+\dots ~,\label{Eq:FieldExpansion1stOrder} \\
  T^I= T^I_{(0)}+ T^I_{(1)}+\dots ~. \label{Eq:FieldExpansion1stOrder_T}
 \end{align}
 where it is assumed that the remainder is $\mathcal{O}(\theta^2)$.
 Recalling the expansions (\ref{Eq:ExpansionK}), (\ref{Eq:ExpansionB}), and using the perturbative expansion (\ref{Eq:FieldExpansion1stOrder}) we obtain
 \begin{align}
 B^{IJ}&= e^I\wedge e^J-\tilde{e}^I_{(0)}\wedge \tilde{e}^J_{(0)}-2\tilde{e}^{[I}_{(0)}\wedge \tilde{e}^{J]}_{(1)}+\mathcal{O}(\theta^2)~, \label{Eq:B_PertuExp}\\
 K^I_{\pha I}&=-2e_{I}\wedge\tilde{e}^{I}_{(0)}-2e_{I}\wedge\tilde{e}^{I}_{(1)}+\frac{i}{2}\theta^{\alpha\beta}\left(\mathcal{L}_{X_\alpha}e^{I}\wedge\mathcal{L}_{X_\beta}e_{I}-\mathcal{L}_{X_\alpha}\tilde{e}^{I}_{(0)}\wedge\mathcal{L}_{X_\beta}\tilde{e}_{(0)I}\right) ~.
 \end{align}
Thus, to first order in $\theta$ the equation of motion (\ref{Eq:WedgeTetradsEquationSecondOrderThetaExpand}) reads as
 \be\label{Eq:WedgeTetradsEquationSecondOrderThetaExpand_FieldsExpand}
 \begin{split}
 \de \left[ -2e_{I}\wedge\tilde{e}^{I}_{(0)}-2e_{I}\wedge\tilde{e}^{I}_{(1)}+\frac{i}{2}\theta^{\alpha\beta}\left(\mathcal{L}_{X_\alpha}e^{I}\wedge\mathcal{L}_{X_\beta}e_{I}-\mathcal{L}_{X_\alpha}\tilde{e}^{I}_{(0)}\wedge\mathcal{L}_{X_\beta}\tilde{e}_{(0)I}\right)\right]+\\
 -4\theta^{\alpha\beta}\mathcal{L}_{X_\alpha}\omega_{+}\wedge \mathcal{L}_{X_\beta}(e_I\wedge \tilde{e}^I_{(0)})-\frac{i}{2}\theta^{\alpha\beta}\mathcal{L}_{X_\alpha}(\omega_+)_{IJ}\wedge \mathcal{L}_{X_\beta}\Big(e^I\wedge e^J-\tilde{e}^I_{(0)}\wedge \tilde{e}^J_{(0)}\Big)=0 ~,
 \end{split}
 \ee
Similarly, from Eq.~(\ref{Eq:NCcaseConnectionEquationSecondOrderThetaExpand}) we obtain the equation for the spin connection $\omega^{IJ}_+$
\be\label{Eq:NCcaseConnectionEquationSecondOrderThetaExpand2} 
\de B^{IJ}+\omega_{+ }^{JK}\wedge B^{I}_{\pha K}-B_K^{\pha J}\wedge\omega_{+ }^{KI}-\frac{i}{2}\theta^{\alpha\beta}\Big(\mathcal{L}_{X_\alpha}\omega_{+ }^{IJ}\wedge \mathcal{L}_{X_\beta}(e_L\wedge \tilde{e}^L_{(0)})\Big)+ \theta^{\alpha\beta}\Big(\mathcal{L}_{X_\alpha}\omega_{+}\wedge \mathcal{L}_{X_\beta}B^{IJ}\Big)=0~,
\ee
which can be recast in a more compact form as
\be\label{Eq:FirstOrder_PerturbationTheory_B}
\de_{\omega} B^{IJ}-\frac{i}{2}\theta^{\alpha\beta}\Big(\mathcal{L}_{X_\alpha}\omega_{+ }^{IJ}\wedge \mathcal{L}_{X_\beta}(e_L\wedge \tilde{e}^L_{(0)})\Big)+ \theta^{\alpha\beta}\Big(\mathcal{L}_{X_\alpha}\omega_{+}\wedge \mathcal{L}_{X_\beta}B^{IJ}\Big)=0~,
\ee
with $B^{IJ}$ given by Eq.~(\ref{Eq:B_PertuExp}) and its covariant derivative defined as
\be
\de_{\omega} B^{IJ}\coloneqq \de B^{IJ}+\omega_{+ }^{JK}\wedge B^{I}_{\pha K}-B_K^{\pha J}\wedge\omega_{+ }^{KI} ~.
\ee

\subsection{A simple example: conformally related unperturbed tetrads}
 We shall apply our perturbative scheme to a particularly simple case, in which the two tetrads are conformally related\footnote{Recall the more general relation given by the no-shear condition (\ref{Eq:GL(2,C)ansatz_commutative}). Equation (\ref{Eq:ConformallyRelatedTetrads}) then corresponds to a trivial Lorentz transformation $\Lambda^I_{\pha J}=\delta^I_{\pha J}$.}
 \be\label{Eq:ConformallyRelatedTetrads}
 \tilde{e}^I_{(0)}= \Omega\,  e^I ~.
 \ee
This is achieved when the one-form $p$ introduced in Eq.~(\ref{Eq:WedgeTetradsExact}) is closed, i.e.~$\de p=0$, see Eqs.~(\ref{Eq:ProjectTildeTetrad}), (\ref{Eq:RelationM_lambda}). Moreover, we assume that $\Omega$ is constant and $\Omega\neq 1$.
  Thus, from the analysis of the commutative case given in Section \ref{Sec:ConnectionEq}, we have  $T^I_{(0)}=0$.
 Therefore, equation (\ref{Eq:WedgeTetradsEquationSecondOrderThetaExpand_FieldsExpand}) implies in this case 
 \be\label{Eq:TetradPerturbativeCorrection_Example}
 \de \left[ -2e_{I}\wedge\tilde{e}^{I}_{(1)}+\frac{i}{2}\theta^{\alpha\beta}(1-\Omega^2)\mathcal{L}_{X_\alpha}e^{I}\wedge\mathcal{L}_{X_\beta}e_{I}\right]-
 \frac{i}{2}(1-\Omega^2)\theta^{\alpha\beta}\mathcal{L}_{X_\alpha}(\omega_+)_{IJ}\wedge \mathcal{L}_{X_\beta}(e^I\wedge e^J)=0
 \ee
 Taking the exterior derivative of Eq.~(\ref{Eq:TetradPerturbativeCorrection_Example}) we have
 \be
 \de\left[ \theta^{\alpha\beta}\mathcal{L}_{X_\alpha}(\omega_+)_{IJ}\wedge \mathcal{L}_{X_\beta}(e^I\wedge e^J) \right]=0 ~,
 \ee
 which implies that the three-form in bracket is locally exact, i.e.~there exists a two-form $q$ such that
 \be\label{Eq:IntroduceFormQ}
 \theta^{\alpha\beta}\mathcal{L}_{X_\alpha}(\omega_+)_{IJ}\wedge \mathcal{L}_{X_\beta}(e^I\wedge e^J)=\de q ~.
 \ee
 Substituting this expression back into Eq.~(\ref{Eq:TetradPerturbativeCorrection_Example}) we have
 \be\label{Eq:TetradPerturbativeCorrection_Example_Recast}
 \de \left[ -2e_{I}\wedge\tilde{e}^{I}_{(1)}+\frac{i}{2}\theta^{\alpha\beta}(1-\Omega^2)\mathcal{L}_{X_\alpha}e^{I}\wedge\mathcal{L}_{X_\beta}e_{I}-\frac{i}{2}(1-\Omega^2) q\right]=0 ~.
 \ee
 Equation (\ref{Eq:TetradPerturbativeCorrection_Example_Recast}) locally implies the existence of a one-form $\tau$ such that
 \be
 -2e_{I}\wedge\tilde{e}^{I}_{(1)}+\frac{i}{2}\theta^{\alpha\beta}(1-\Omega^2)\mathcal{L}_{X_\alpha}e^{I}\wedge\mathcal{L}_{X_\beta}e_{I}-\frac{i}{2}(1-\Omega^2) q=\de\tau ~.
 \ee
 Thus, in this simple example the first order perturbative correction to the tilde tetrad is given by the solution of the following algebraic equation
  \be\label{Eq:TetradPerturbativeCorrection_Solution}
  e_{I}\wedge\tilde{e}^{I}_{(1)}=\frac{i}{4}\theta^{\alpha\beta}(1-\Omega^2)\mathcal{L}_{X_\alpha}e^{I}\wedge\mathcal{L}_{X_\beta}e_{I}-\frac{i}{4}(1-\Omega^2) q-\frac{1}{2}\de\tau  ~.
 \ee
 The one-form $\tau$ in Eq.~(\ref{Eq:TetradPerturbativeCorrection_Solution}) is entirely arbitrary and its contribution is not related to spacetime noncommutativity.
  Thus, we can set $\de\tau=0$ in order to single out the effects of noncommutativity in determining the corrections to $\tilde{e}^{I}$. The two-form $q$ is instead obtained by solving the differential equation (\ref{Eq:IntroduceFormQ}), which admits local solutions for $q$ provided the l.h.s.~is a closed three-form. Thus, Eq.~(\ref{Eq:IntroduceFormQ}) also constrains the functional form of $(\omega_+)_{IJ}$ and $e^I$. Note that in general the second term in Eq.~(\ref{Eq:TetradPerturbativeCorrection_Solution}) is non-vanishing even when the spin connection $(\omega_+)_{IJ}$ is pure gauge.

The equation of motion for the connection (\ref{Eq:FirstOrder_PerturbationTheory_B}) becomes
 \be\label{Eq:TildeTorsionExpand_Perturbative}
\de_{\omega} B^{IJ}+\theta^{\alpha\beta} (1-\Omega)^2\mathcal{L}_{X_\alpha}\omega_{+}\wedge \mathcal{L}_{X_\beta}(e^I\wedge e^J)=0~,
\ee
 which, using Eq.~(\ref{Eq:B_PertuExp}), can be recast in the form
 \be\label{Eq:ConnectionEq_Pert_SecondToLast}
 2(1-\Omega^2)e^{[I}\wedge T^{J]}_{(1)}+\theta^{\alpha\beta} (1-\Omega)^2\mathcal{L}_{X_\alpha}\omega_{+}\wedge \mathcal{L}_{X_\beta}(e^I\wedge e^J)=0 ~.
 \ee
  Thus, we have the equation determining the first order perturbative corrections to the torsion
   \be\label{Eq:Torsion_PerturbativeSolution}
 e^{[I}\wedge T^{J]}_{(1)}=-\frac{1}{2}\theta^{\alpha\beta} \mathcal{L}_{X_\alpha}\omega_{+}\wedge \mathcal{L}_{X_\beta}(e^I\wedge e^J) ~.
 \ee
 Equation (\ref{Eq:Torsion_PerturbativeSolution}) shows that, already in this simple example, torsion is sourced by spacetime noncommutativity, provided that $\omega_{+}$ is not a constant. The perturbative corrections are first order in $\theta$.

 \section{Further extensions of the model}\label{interact}
 \subsection{Bitetrad interactions}
It is possible to include the following interaction terms in the action
\be\label{Eq:InteractionTermBimetric}
S_{\rm int}[e^I, \tilde{e}^I]=\int \Big\{ i\,c_1\Tr\left[{\bm e}\wedge_{\star}{\bm e}\wedge_{\star}{\bm e}\wedge_{\star}{\bm e}\,\gamma_5\right] +c_2 \Tr\left[{\bm e}\wedge_{\star}{\bm e}\wedge_{\star}{\bm e}\wedge_{\star}{\bm e}\right] +c_3 \left(\Tr\left[{\bm e}\wedge_{\star}{\bm e}\right]\right)^2+c_4\left(\Tr\left[{\bm e}\wedge_{\star}{\bm e}\,\gamma_5\right]\right)^2\Big\}
\ee
These are the simplest (polynomial) interaction terms in four dimensions that are compatible with the symmetries of the model, and that do not give rise to higher-order derivatives in the commutative limit. For generality, in this section we will not make use of the no-shear condition (\ref{Eq:GL(2,C)ansatz_commutative}), although it will be pointed out when simplifications arise due to such an assumption. It is straightforward, if tedious, to expand $S_{\rm int}$ by explicitly evaluating the traces and using the graded ciclity property of the $\wedge_{\star}$ product (see Eq.~(\ref{Eq:GradedCiclicity})). Denoting by $S_{i}$ the term multiplying the coefficient $c_i$ in Eq.~(\ref{Eq:InteractionTermBimetric}), we obtain
\begin{align}
S_1 &=4\,\epsilon_{IJKL}\int\left(e^I\wedge_{\star}e^J\wedge_{\star}e^K\wedge_{\star}e^L+\tilde{e}^I\wedge_{\star}\tilde{e}^J\wedge_{\star}\tilde{e}^K\wedge_{\star}\tilde{e}^L\right)+\\
&\phantom{=} 
-8\,\epsilon_{IJKL}\int\left(2 e^I\wedge_{\star}e^J\wedge_{\star}\tilde{e}^K\wedge_{\star}\tilde{e}^L-e^I\wedge_{\star}\tilde{e}^J\wedge_{\star}e^K\wedge_{\star}\tilde{e}^L\right)+\nonumber\\
&\phantom{=} 
16\,i\int\left(e^I\wedge_{\star}e_I\wedge_{\star}e^J\wedge_{\star}\tilde{e}_J-e^I\wedge_{\star}e^J\wedge_{\star}e_I\wedge_{\star}\tilde{e}_J +e^I\wedge_{\star}e^J\wedge_{\star}e_J\wedge_{\star}\tilde{e}_I\right)+ \nonumber\\
&\phantom{=} 
16\,i\int\left(\tilde{e}^I\wedge_{\star}\tilde{e}_I\wedge_{\star}\tilde{e}^J\wedge_{\star}e_J-\tilde{e}^I\wedge_{\star}\tilde{e}^J\wedge_{\star}\tilde{e}_I\wedge_{\star}e_J +\tilde{e}^I\wedge_{\star}\tilde{e}^J\wedge_{\star}\tilde{e}_J\wedge_{\star}e_I\right) \nonumber
\\
S_2 &=-4\int\left(e^I\wedge_{\star}e^J\wedge_{\star}e_I\wedge_{\star}e_J+\tilde{e}^I\wedge_{\star}\tilde{e}^J\wedge_{\star}\tilde{e}_I\wedge_{\star}\tilde{e}_J\right)\\
S_3 &=16\int\left(e^I\wedge_{\star}e_I\wedge_{\star}e^J\wedge_{\star}e_J+\tilde{e}^I\wedge_{\star}\tilde{e}_I\wedge_{\star}\tilde{e}^J\wedge_{\star}\tilde{e}_J-2\, e^I\wedge_{\star}e_I\wedge_{\star}\tilde{e}^J\wedge_{\star}\tilde{e}_J\right)\\
S_4 &=16\int\left(e^I\wedge_{\star}\tilde{e}_I\wedge_{\star}e^J\wedge_{\star}\tilde{e}_J+\tilde{e}^I\wedge_{\star}e_I\wedge_{\star}\tilde{e}^J\wedge_{\star}e_J-2\, e^I\wedge_{\star}\tilde{e}_I\wedge_{\star}\tilde{e}^J\wedge_{\star}e_J\right)
\end{align}
In the commutative limit, many such terms vanish identically. Thus, we have
\begin{align}
S_1^{\scriptscriptstyle\theta=0} &=4\,\epsilon_{IJKL}\int\left(e^I\wedge e^J\wedge e^K\wedge e^L+\tilde{e}^I\wedge \tilde{e}^J\wedge \tilde{e}^K\wedge \tilde{e}^L\right)   \label{Eq:Interaction_S1_theta0}
-8\,\epsilon_{IJKL}\int\left(e^I\wedge e^J\wedge\tilde{e}^K\wedge \tilde{e}^L \right)\nonumber\\
S_2^{\scriptscriptstyle\theta=0} &=0\\
S_3^{\scriptscriptstyle\theta=0} &=0\\
S_4^{\scriptscriptstyle\theta=0} &=64\int e^I\wedge\tilde{e}_I\wedge e^J\wedge\tilde{e}_J \label{Eq:S4_CommutativeLimit}
\end{align}
The term (\ref{Eq:Interaction_S1_theta0}) is of the type of consistent interactions in (tetrad) ghost-free bigravity \cite{hinterbichler}. Note that the tetrad formulation of bimetric gravity is equivalent to their metric formulation provided that the Deser-van Nieuwenhuizen condition holds \cite{hinterbichler,deffayet} (see also Refs.~\cite{tamanini,hoek})
\be\label{Eq:Deser-vanNieuwenhuizen}
e^I\wedge\tilde{e}_I=0 ~.
\ee
The geometric meaning of (\ref{Eq:Deser-vanNieuwenhuizen}) is explained in \cite{Hassan:2017ugh}.
In our model, this condition would also imply that the equation of motion (\ref{Eq:EOM1_Commutative}) is automatically satisfied.
When the condition expressed by Eq.~(\ref{Eq:Deser-vanNieuwenhuizen}) is satified, the vanishing of the interaction term (\ref{Eq:S4_CommutativeLimit}) also follows necessarily.
Therefore, we obtain in this case a one-parameter\footnote{As far as interactions between the two tetrads are concerned.} bigravity model, with interaction term given by
\be\label{Eq:Sint_commutative}
S_{\rm int}^{\scriptscriptstyle\theta=0}=4\,c_1\int \epsilon_{IJKL}\left( e^I\wedge e^J\wedge e^K\wedge e^L+\tilde{e}^I\wedge \tilde{e}^J\wedge \tilde{e}^K\wedge \tilde{e}^L
-2\, e^I\wedge e^J\wedge\tilde{e}^K\wedge \tilde{e}^L \right) ~.
\ee
It is interesting to observe that the interaction term (\ref{Eq:Sint_commutative}) corresponds to the partially massless bigravity theory\footnote{We are thankful to Latham Boyle and Fawad Hassan for pointing out this correspondence.}, first indentified in \cite{Hassan:2012gz}.

\subsection{Higher-order curvature invariants}
There are only two possible monomial invariants that can be built using only the field strength and its dual. The corresponding actions are quadratic in the curvature and read as
\begin{align}
S_{\rm\scriptscriptstyle P}&=\int \Tr\left[{\bm F}\wedge_{\star}{\bm F}\right]=-\frac{1}{2}\int F^{IJ}\wedge_{\star}F_{IJ}+4\int\left(r\wedge_{\star}r+\tilde{r}\wedge_{\star}\tilde{r}\right) ~,\label{Eq:Pontryagin_Action}
\\
S_{\rm\scriptscriptstyle MM}&=\int \Tr\left[{\bm F}\wedge_{\star} *_{\Hdg}{\bm F}\right]=\frac{1}{2}\int F^{IJ}\wedge_{\star} *_{\Hdg}F_{IJ}-4\,i\int\left(r\wedge_{\star}\tilde{r}+r\wedge_{\star}\tilde{r}\right) ~.\label{Eq:MacDowell-Mansouri_Action}
\end{align}
Such action functionals represent the noncommutative extensions of the Pontryagin action and the MacDowell-Mansouri action \cite{macdowell}, respectively. The former reduces to a topological term in the commutative case. The noncommutative extension of the MacDowell-Mansouri action,  Eq.~(\ref{Eq:MacDowell-Mansouri_Action}), has been previously obtained in Ref.~\cite{aschieri}. The MacDowell-Mansouri action was also studied in a bimetric setting in relation to partial masslessness in \cite{Apolo:2016ort}. We observe that, if a self-dual gauge connection is assumed, then there is a simple relation between the actions (\ref{Eq:Pontryagin_Action}), (\ref{Eq:MacDowell-Mansouri_Action}) (see Ref.~\cite{Montesinos:2001ww} for the commutative case). This is analogous to the relation between the Palatini and the Holst term in the self-dual case, which we examined in this paper.

\section{Conclusions}\label{concl}
In this work, we generalized the model of Ref.~\cite{aschieri} and built a noncommutative extension of tetrad Palatini-Holst gravity, based on an Abelian twist. In the framework adopted, the noncommutative deformation necessarily leads to the enlargement of the internal gauge symmetry of the model, which is thus extended from the Lorentz group to $\rm GL(2,\C)$. Similar consistency requirements demand that the metric degrees of freedom of the theory must be also augmented, thus replacing the tetrad by a bitetrad. Therefore, the theory obtained is inherently bimetric. We take the standpoint that the extra degrees of freedom required by the noncommutative extension are physical. Thus, a modified theory of gravity entailing both higher-order derivatives and new gravitational degrees of freedom is obtained, which naturally encodes modifications of spacetime structure at the Planck scale.

The inclusion of the Holst term in the action of the noncommutative theory has important consequences for the dynamics. In fact, by choosing the value $\beta=-i$ for the Barbero-Immirzi parameter\footnote{We recall that $\beta$ is the inverse of the coupling of the Holst term.}, the action turns out to depend only on the self-dual part of the $\frak{gl}(2,\C)$ gauge connection. This result is a generalization of a well-known property of the corresponding commutative theory. Moreover, in the noncommutative case, the field strength has a much simpler expression in the self-dual case and its components are given in Eqs.~(\ref{Eq:Eq:FieldStrenghtComponents1}), (\ref{Eq:Eq:FieldStrenghtComponents2}). This in turn leads to a great simplification in the equations of motion compared to the general case. The equations of motion for the self-dual action are given in Eqs.~(\ref{Eq:NCcaseFieldEquation1}), (\ref{Eq:NCcaseFieldEquation2}), (\ref{Eq:NCcaseWedgeTetradsEquation}), (\ref{Eq:NCcaseConnectionEquation}).

In Section~\ref{symm} we studied the symmetries of the model for generic $\beta$. These are of two types, namely gauge symmetries and duality symmetries. In the first class, we have spacetime symmetries (diffeomorphisms, $\star$-diffeomorphisms), as well as the internal $\rm GL(2,\C)$ gauge $\star$-symmetry. In the second class, we identified three distinct duality symmetries in the target space. One of them is a straightforward generalization of Hodge duality, while the other two rest on the doubling of the tetrad degrees of freedom required by the noncommutative deformation.

The commutative limit of the theory was studied in detail in Section~\ref{commlim}. In particular, we showed that the dynamics imposes a constraint on the relation between the two tetrads (\ref{Eq:WedgeTetradsClosed}), and determined solutions to the generalized connection equation (\ref{EQ:eomCommutative}). Remarkably, torsionful connections are admissible solutions even in the pure gravity case. The extra component in the $\frak{gl}(2,\C)$ gauge connection shall instead be identified with the Weyl one-form, representing a particular type of spacetime non-metricity. Thus, the commutative limit of the theory turns out to be invariant under local scale transformations. Equations~(\ref{Eq:EinsteinU_CommutativeCase_R1}), (\ref{Eq:EinsteinV_CommutativeCase_R1}) give the gravitational field equations for the two tetrads, in differential forms notation; they can be recast in the equivalent tensorial form (\ref{Eq:FieldEquations_FinalForm}). In Section~\ref{Sec:Reality}, we imposed reality conditions on physical fields. Thus, we showed that the form of the gravitational field equations is similar to Einstein-Cartan theory, with the field strength of the Weyl non-metricity one-form acting as a source of torsion. Therefore, torsion would be dynamical even in vacuo, in contrast to standard Einstein-Cartan theory. 

In Section~\ref{thetaexp}, we study the effects of spacetime noncommutativity by adopting a perturbative approach. More specifically, by means of an asymptotic expansion of the twist operator in the deformation parameter $\theta$, we obtain correction terms to the equations of motion of the commutative theory. Then, focusing on the connection equation and the bitetrad constraint, we determine corrections to solutions of the commutative theory. We consider as a simple example the unperturbed solution corresponding to two tetrads related by a constant scale transformation, and vanishing torsion. We show that even in this simple case there are non-trivial perturbative corrections to first order in $\theta$; in particular, torsion is shown to receive non-vanishing corrections.

Lastly, in Section~\ref{interact}, we consider further extensions of the model. Particularly interesting in this regard are extensions achieved by the inclusion of self-interaction terms of the bitetrad. We showed that there are only four possible such terms that are polynomial and compatible with the gauge symmetries of the model. In the commutative limit, they give rise to interaction terms that are typical of ghost-free bimetric theories of gravity, see Eq.~(\ref{Eq:Sint_commutative}). It is worth noting that, in the $\theta\to 0$ limit, there is only one free parameter in the interaction term.

\newpage

\appendix

\section{Elements of twisted differential geometry}\label{noncommapp}
Twist differential geometry is a powerful tool that allows to construct noncommutative spacetimes as deformations of commutative spacetimes, while retaining associativity. The noncommutative structure underlying the model studied in the present work is the one introduced in Ref.~\cite{aschieri}; it is  obtained by means of a particular type of Abelian twist, which is used to generalize the Moyal-Weyl $\star$-product to the exterior algebra of differential forms on a spacetime manifold. Let us shortly review the setup, while referring to \cite{aschieri} and references therein for further details. The twist will be denoted by $\mathcal{F}\in U\Xi \otimes U\Xi $, where  $U\Xi$  is the universal enveloping algebra of the Lie algebra $\Xi$ of smooth tensor fields on spacetime\footnote{Products of Lie algebra elements are generally not Lie algebra elements themselves. The universal enveloping algebra is an algebra whose elements are all such products. It is infinite-dimensional and contains all representations of the given Lie algebra.}. Using the twist $\mathcal{F}$, a $\star$-product of smooth functions $f, g\in C^\infty(M)$ can be defined as a deformation of the ordinary pointwise multiplication
\be\label{Eq:MoyalWeylProduct}
f\star g=\mu\circ \mathcal{F}^{-1}(f\otimes g)  ~, 
\ee
which can be easily generalized to fields with non-zero spin. 
The bilinear operator $\mu$ denotes pointwise multiplication, i.e.
\be
\mu (f\otimes g) =f\cdot g ~.
\ee
We will assume a twist of the form
\be\label{Eq:TwistDefinition_1}
\mathcal{F}=\e^{ -\frac{i}{2}\theta^{\alpha\beta}X_\alpha\otimes X_\beta } ~,
\ee
where $\{X_\alpha\}$ is a set of mutually commuting vector fields (Abelian twist). The matrix $\theta^{\alpha\beta}$ is assumed constant and antisymmetric
\be
\theta^{\alpha\beta}=-\theta^{\beta\alpha} ~.
\ee
  If we choose a system of local coordinates $\{x^\alpha\}$ adapted to the vector fields $\{X_\alpha\}$, i.e.~such that $\frac{\pa\phantom{f}}{\pa x^{\alpha}}=X_\alpha$, the $\star$-product defined in Eq.~(\ref{Eq:MoyalWeylProduct}) reduces to the usual definition of the Moyal-Weyl product
\be
f\star g=\mu\circ \e^{ \frac{i}{2}\theta^{\alpha\beta}\pa_\alpha\otimes \pa_\beta } (f\otimes g)  ~.
\ee

The generalization of the definition (\ref{Eq:MoyalWeylProduct}) to the twist deformation of more general bilinear composition laws, such as e.g., the tensor product $\otimes$ of smooth tensor fields and the wedge product $\wedge$ of smooth differential forms, is readily obtained from Eq.~(\ref{Eq:TwistDefinition_1}) by replacing the vector field $X_\alpha$ with the corresponding Lie derivative $\mathcal{L}_{X_\alpha}$
\be\label{Eq:TwistDefinition_2}
\mathcal{F}=\e^{ -\frac{i}{2}\theta^{\alpha\beta}\mathcal{L}_{X_\alpha}\otimes \mathcal{L}_{X_\beta} } ~.
\ee
Moreover, we find it convenient to use the standard notation
\be
\mathcal{F}=\mathcal{F}^\alpha\otimes\mathcal{F}_\alpha,~~~~\mathcal{F}^{-1}=\bar{\mathcal{F}}^\alpha\otimes\bar{\mathcal{F}}_\alpha
\ee
with $\mathcal{F}^\alpha$ multi-differential operators and $\alpha$ a collective index.

Thus, considering two tensor fields $s$, $t$, we can define the twist-deformation of their tensor product as
\be
s \otimes_\star t= \bar{\mathcal{F}}^\alpha(s)\otimes \bar{\mathcal{F}}_\alpha(t)  ~,
\ee
Similarly, the $\star$-deformed wedge product of two differential forms is defined as
\be\label{Eq:DeformedWedgeProd_Define}
 \xi \wedge_\star \eta= \bar{\mathcal{F}}^\alpha(\xi)\wedge \bar{\mathcal{F}}_\alpha(\eta) .
\ee

The deformed wedge product inherits some of the properties of its commutative counterpart:
\begin{enumerate}[label=\roman*)]
\item It is associative, i.e.~given three differential forms $\xi$, $\eta$, $\tau$, we have
\be
(\xi\wedge_{\star}\eta)\wedge_{\star}\tau=\xi\wedge_{\star}(\eta\wedge_{\star}\tau) ~.
\ee
\item The $\wedge_{\star}$ product of a differential form of arbitrary degree $\xi$ and a 0-form $f$ (i.e.~a scalar function) reduces to the ordinary $\star$-product
\begin{align}
f\wedge_{\star} \xi=f\star \xi \label{Eq:CompatibilityWedgeStar_StarProd1}\\
\xi\wedge_{\star} f=\xi\star f \label{Eq:CompatibilityWedgeStar_StarProd2}
\end{align}
\item The action of the exterior derivative is compatible with the twist, i.e.~a graded Leibniz rule holds
\be\label{Eq:GradedCiclicity}
\de (\sigma \wedge_{\star}\tau)= \de \sigma \wedge_{\star}\tau + (-1)^{\rm deg(\sigma)}\sigma\wedge_{\star}\de\tau
\ee
\item The deformed wedge product satisfies a graded ciclicity property
\be\label{Eq:GradedCiclicity}
\int \sigma \wedge_{\star}\tau = (-1)^{\rm deg(\sigma)deg(\tau)}\int \tau \wedge_{\star}\sigma ~,
\ee
where ${\rm deg}(\sigma)+{\rm deg}(\tau)=D$, $D$ being the number of spacetime dimensions, and the equality holds up to boundary terms.
\item Compatibility with undeformed complex conjugation
\be\label{Eq:ComplexConjugation}
\overline{(\sigma \wedge_{\star}\tau)}=(-1)^{\rm deg(\sigma)deg(\tau)}\, \overline{\tau}\wedge_{\star}\overline{\sigma} ~.
\ee
In the Moyal-Weyl case and for real forms, complex conjugation of the wedge product on the l.h.s.~of Eq.~(\ref{Eq:ComplexConjugation}) is equivalent to the sign reversal of $\theta^{\alpha\beta}$; thus implying
\be
\sigma \wedge_{\star_{-\theta}}\tau=(-1)^{\rm deg(\sigma)deg(\tau)}\, \tau\wedge_{\star_{\theta}}\sigma  ~.
\ee
\end{enumerate}

Finally, it is convenient for perturbative computations to express the $\wedge_{\star}$ product as a series expansion in the parameters $\theta^{\alpha\beta}$. From the definitions (\ref{Eq:DeformedWedgeProd_Define}), (\ref{Eq:TwistDefinition_2}) we find, after expanding the exponential in the definition of the twist
\be\label{Eq:WedgeStar_Expansion}
\xi \wedge_\star \eta=\xi \wedge \eta+\frac{i}{2}\theta^{\alpha\beta}\mathcal{L}_{X_\alpha}\xi\wedge\mathcal{L}_{X_\beta}\eta+\frac{1}{2!}\left(\frac{i}{2}\right)^2\theta^{\alpha\beta}\theta^{\rho\sigma}\mathcal{L}_{X_\rho}\mathcal{L}_{X_\alpha}\xi\wedge\mathcal{L}_{X_\sigma}\mathcal{L}_{X_\beta}\eta +\mathcal{O}(\theta^3) ~.
\ee
We remark that, unlike the ordinary wedge product, the $\wedge_\star$ product fails to satisfy a graded anticommutativity property. In fact, it is clear from the expansion (\ref{Eq:WedgeStar_Expansion}) that, due to the $\theta$-dependend corrections, one has $\xi \wedge_\star \eta \neq (-1)^{\rm deg(\xi)deg(\eta)} \eta\wedge_\star\xi$ in general.

\section{Useful formulae involving Dirac Gamma matrices}\label{Clifford}
\be
\{\gamma_I,\gamma_J\}=2\eta_{IJ}~,~~~~\eta_{IJ}=(+,-,-,-)
\ee
\be
\gamma^I=\eta^{IJ}\gamma_J=\gamma_0\gamma_I\gamma_0
\ee
\be
\gamma_5\coloneqq i\gamma_0\gamma_1\gamma_2\gamma_3=-\frac{i}{4!}\varepsilon^{IJKL}\gamma_{I}\gamma_{J}\gamma_{K}\gamma_{L}
\ee
\be
\gamma_{I}\gamma_{J}\gamma_{K}\gamma_{L}=-i\varepsilon_{IJKL}\gamma_5
\ee
\be
\gamma_I\gamma_J\gamma_K=\eta_{IJ}\gamma_K-\eta_{IK}\gamma_J+\eta_{JK}\gamma_I+i\varepsilon_{IJKL}\gamma^L\gamma_5
\ee
\be
\Gamma_{IJ}=\frac{i}{4}[\gamma_I,\gamma_J]~
\ee
\be
\Gamma_{IJ}\gamma_5=\frac{i}{2}\varepsilon_{IJKL}\Gamma^{KL}
\ee
\be
P_{\pm}=\frac{1\mp \gamma_5}{2}
\ee
\be
\Gamma_{IJ}P_{\pm}=\frac{1}{2}\left(\Gamma_{IJ}\mp\frac{i}{2}\varepsilon_{IJKL}\Gamma^{KL}\right)=p^{\pm}_{IJKL}\Gamma^{KL} ~.
\ee
\be\label{Appendix:Eq:GeneratorsCommutator}
[\Gamma_{IJ},\Gamma_{KL}]=i\left(\eta_{IL}\Gamma_{JK}-\eta_{IK}\Gamma_{JL}+\eta_{JK}\Gamma_{IL}-\eta_{JL}\Gamma_{IK}\right)
\ee
\be\label{Appendix:Eq:GeneratorsAntiCommutator}
\{\Gamma_{IJ},\Gamma_{KL}\}=
\eta_{I[K}\eta_{L]J}+\frac{i}{2}\varepsilon_{IJKL}\gamma_5
\ee
\be
[\gamma_K,\Gamma_{IJ}]=i(\eta_{KI}\gamma_J-\eta_{KJ}\gamma_I)
\ee
\be
[\gamma_5,\Gamma_{IJ}]=0
\ee
\be
\{\gamma_I,\Gamma_{JK}\}=-\varepsilon_{IJKL}\gamma^L \gamma_5
\ee
\be
\Tr\left[\gamma_I\gamma_J\right]=4\eta_{IJ}
\ee
\be
\Tr\left[\gamma_I\gamma_J\gamma_5\right]=0
\ee
\be
\Tr\left[\gamma_I\gamma_J\gamma_K\gamma_L\right]=4\left(\eta_{IJ}\eta_{KL}-\eta_{IK}\eta_{JL}+\eta_{IL}\eta_{JK}\right)
\ee
\be
\Tr\left[\gamma_I\gamma_J\gamma_K\gamma_L\gamma_5\right]=-4i\varepsilon_{IJKL}
\ee
\be
\Tr\left[\gamma_I\gamma_J\Gamma_{KL}\right]=
-4i\,\eta_{I[L}\eta_{K]J}
\ee
\be
\Tr\left[\gamma_I\gamma_J\Gamma_{KL}\gamma_5\right]=2\varepsilon_{IJKL}
\ee
\section{Tetrad identities}
\be
g^{ab} e_{a}^I e_b^J=\eta^{IJ}
\ee
\be
\eta_{IJ} e_a^I e_b^J=g_{ab}
\ee
\be
e\coloneqq \det e=\sqrt{-\det g}
\ee
The Levi-Civita symbol is defined so as to satisfy the conventions $\varepsilon^{0123}=-1$ and $\varepsilon_{0123}=1$.
\be
\varepsilon^{abcd}e_a^I e_b^J e_c^K e_d^L= \varepsilon^{IJKL}\, e
\ee
\be
\varepsilon_{IJKL}\varepsilon^{abcd}e_a^I e_b^J e_c^K e_d^L=- 4!\, e
\ee
\be
\varepsilon^{abcd}e_a^I e_b^J=\varepsilon^{IJKL} e\, e^c_K e^d_L
\ee
\be
\varepsilon_{IJKL}\varepsilon^{abcd} e_a^I e_b^J=-4 e\, e^c_{[K}e^d_{L]}
\ee

\section{Weyl connection}\label{Appendix:Weyl}
The Weyl connection can be introduced by means of the following compatibility condition\footnote{Sometimes this condition is improperly referred to as `tetrad postulate'.} between the affine connection (specified by the assignment of the connection coefficients $\Gamma_{ab}^c$) and the spin connection (specified by $\omega_{a\pha J}^{\pha I}$ and $w_{a}$)
\be\label{Eq:TetradPostulate}
\nabla_{a}e_b^I+\omega_{a\pha J}^{\pha I}e^J_{\pha b}-\frac{1}{2}w_{a} e^I_{b}=0
\ee
Antisymmetrising over the pair of indices $(a,b)$ and using differential forms notation, we obtain from Eq.~(\ref{Eq:TetradPostulate})
\be
\de e^I + \omega^I_{\pha J}\wedge e^J - \frac{1}{2} w\wedge e^I = C^I ~,
\ee
where $C^I_{ab}=\Gamma_{[ab]}^I$.
This relation can be written more compactly by introducing a new spin connection, which includes a 
contribution that is symmetric in the internal space
\be\label{Eq:NewConnection}
 \bar{\omega}^{IJ}\coloneqq \omega^{IJ}-\frac{1}{2}\eta^{IJ} w ~.
 \ee
The Weyl vector introduces a particular type of non-metricity (pure trace). In fact, we have
\be
Q_{IJ} \coloneqq \de_{\bar{\omega}} \eta_{IJ}= \de \eta_{IJ} - \bar{\omega}_I^{\pha K} \eta_{KJ}-\bar{\omega}_J^{\pha K} \eta_{IK}=-\bar{\omega}_{IJ}-\bar{\omega}_{JI}=\eta_{IJ} w ~.
\ee

The curvature of the connection (\ref{Eq:NewConnection}) is defined as
\be\label{Eq:BarredFieldStrength_Definition}
\bar{F}^{IJ}(\bar{\omega})=\de\bar{\omega}^{IJ}+\bar{\omega}^I_{\pha K}\wedge \bar{\omega}^{KJ}~.
\ee
The quantity $\bar{F}^{IJ}(\bar{\omega})$ is both Lorentz and Weyl invariant. A straightforward calculation gives
\be\label{Eq:BarredFieldStrength_Expression}
\bar{F}^{IJ}(\bar{\omega})=F^{IJ}(\omega)-\frac{1}{2}\eta^{IJ}\de w ~.
\ee

\section{Torsionful Geometry}\label{Appendix:Torsion}
In this Appendix we review the geometry of a spacetime with curvature and torsion\footnote{The reader must be aware of some differences between our conventions and those adopted in the above references. For instance, the relation between our definition of the torsion tensor and the one used in Ref.~\mciteSubRef{Hehl:1976kj} is $T^{c}_{\pha ab}=2S_{ab}^{\pha\pha c}$.}
 \cite{\torsion,capozziello-torsion}. Let us consider a torsionful affine connection $\nabla_a$. The torsion tensor is defined by the following relation
\be
[\nabla_a,\nabla_b]f=-T^{c}_{\pha ab}\nabla_c f ~,
\ee
where $f$ is a scalar function.
The connection $\nabla_a$ is assumed to be metric compatible, i.e.~it satisfies
\be
\nabla_{c}\, g_{ab}=0 ~.
\ee
The corresponding connection coefficients  are
\be
\Gamma^c_{\pha ab}= \overline{\Gamma}^c_{\pha ab}+K^c_{\pha ab} ~,
\ee
where the first term is a Christoffel symbol and $K^c_{\pha ab}$ is the contortion tensor, defined as
\be
K^c_{\pha ab}\coloneqq \frac{1}{2}\left( T^c_{\pha ab}+T^{\;c}_{a\pha b}+T^{\;c}_{b\pha a}\right) ~.
\ee
Thus, we have
\be
\Gamma^c_{\pha[ab]}=K^c_{\pha[ab]}=\frac{1}{2}T^c_{\pha ab} ~.
\ee
With our conventions, the contortion tensor is antisymmetric w.r.t.~to its first and third indices
\be\label{Eq:ContortionAntisymmetry}
K_{abc}=-K_{c b a} ~.
\ee
It is convenient to define the following contraction of the torsion tensor
\be
 T_a\eqqcolon T^{c}_{\pha ac} ~.
 \ee
 The contraction of the contortion tensor over its first and third indices vanishes due to Eq.~(\ref{Eq:ContortionAntisymmetry})
\be
K^c_{\pha ac}=0 ~.
\ee
We also have
\be
K^a_{\pha a b}=-T_b ~,
\ee
while contraction with the inverse metric gives
\be
g^{ab}K_{cab}= T_c ~.
\ee

The Riemann tensor of a torsionful connection can be defined through its action on a vector field $V^a$ as
\be
R_{abc}^{\pha\pha\pha d}V^c=(\nabla_a\nabla_b-\nabla_b\nabla_a)V^d+T^c_{\pha ab}\nabla_c V^d ~.
\ee
For a one-form $\omega_a$ we have instead
\be
-R_{abc}^{\pha\pha\pha d}\omega_d=(\nabla_a\nabla_b-\nabla_b\nabla_a)\omega_c+T^e_{\pha ab}\nabla_e \omega_c 
\ee
The Riemann curvature tensor of the torsionful connection $\nabla_a$ can be expanded in terms of the curvature of the metric connection and terms involving the torsion.
 An overline is used to denote the Levi-Civita connection and the corresponding curvature tensor, as well as its contractions (cf.~e.g.~Ref.~\cite{capozziello-torsion})
\be
R_{abc}^{\pha\pha\pha d}=\overline{R}_{abc}^{\pha\pha\pha d}+\overline{\nabla}_a K^d_{\pha bc}-\overline{\nabla}_b K^d_{\pha ac}+K^d_{\pha a e}K^e_{\pha bc}-K^d_{\pha b e}K^e_{\pha ac}
\ee
\be
\begin{split}
R_{ab}\coloneqq R_{cab}^{\pha\pha\pha c}=&\overline{R}_{ab}+\overline{\nabla}_c K^c_{\pha ab}+\overline{\nabla}_a T_{b}-T_c K^c_{\pha ab}-K^c_{\pha a e}K^e_{\pha c b}  \label{Eq:RicciTensorWithTorsion}
\end{split}
\ee
\be
R\coloneqq g^{ab}R_{ab}=
\overline{R}+2\overline{\nabla}_a T^a-T^a T_a -K_{abc}K^{cab}  ~. \label{Eq:RicciScalarWithTorsion}
\ee
The Einstein tensor is defined as usual
\be
G_{ab}\coloneqq R_{ab}-\frac{1}{2}g_{ab} R ~,
\ee
with the Ricci tensor $R_{ab}$ and the Ricci scalar defined by Eqs.~(\ref{Eq:RicciTensorWithTorsion}) and (\ref{Eq:RicciScalarWithTorsion}), respectively. Note that unlike in general relativity the Einstein tensor $G_{ab}$ is not symmetric. Lastly, the Bianchi identities in a spacetime endowed with torsion read as (see Ref.~\cite{hehl-datta})
\begin{align}
&R_{[abc]}^{\pha\pha\pha\pha d}=\nabla_{[a}T^d_{\pha bc]}-T^e_{\pha [ab}T^d_{\pha c] e} ~,\label{Eq:Bianchi1} \\
&\nabla_{[a} R_{bc]d}^{\pha\pha\pha e}=T^l_{\pha [ab}R_{c]ld}^{\pha\pha\pha e} ~.\label{Eq:Bianchi2}
\end{align}

\noindent{\bf Acknowledgements} 
The work of M.d.C. is partially supported by the Atlantic Association for Research in the Mathematical Sciences (AARMS), and by a STSM Grant from COST (European Cooperation in Science  and  Technology)  in  the  framework  of  COST  Action  MP1405 ``Quantum Structure of Spacetime''. M.S. and P.V. also  acknowledge partial support from  COST  Action  MP1405 for their work. MdC is thankful for the warm hospitality received at the University of Naples ``Federico II'', where part of this work was done. The authors would like to thank Paolo Aschieri, Leonardo Castellani and Marija Dimitrijevi\'c \'Ciri\'c for discussions. We are thankful to Fawad Hassan for correspondence. M.d.C. would also like to thank Latham Boyle and Viqar Husain for useful comments.
 
 \bibliographystyle{apsrev4-1M}
 \bibliography{nc_holst}

\end{document}